\documentclass{article}

\usepackage{PRIMEarxiv}

\usepackage[utf8]{inputenc} 
\usepackage[T1]{fontenc}    
\usepackage{hyperref}       
\usepackage{url}            
\usepackage{booktabs}       
\usepackage{amsfonts}       
\usepackage{amsmath,bm}
\usepackage{nicefrac}       
\usepackage{microtype}      
\usepackage{lipsum}
\usepackage{fancyhdr}       
\usepackage{graphicx}       
\usepackage[font=small,labelfont=bf]{caption}
\usepackage{subfigure}
\usepackage{xcolor}
\usepackage{natbib}

\newcommand{\p}{\mathbb{P}}
\newcommand{\E}{\mathbb{E}}
\newcommand{\Var}{\text{Var}}
\newcommand{\bd}[1]{\boldsymbol{#1}}

\title{The hidden factor: accounting for covariate effects in power and sample size computation for a binary trait}

\author{
  Ziang Zhang \\
  Department of Statistical Science \\
  University of Toronto \\
  Toronto\\
  \texttt{aguero.zhang@mail.utoronto.ca} \\
   \And
  Lei Sun \\
  Department of Statistical Science \\
  University of Toronto \\
  Toronto\\
  \texttt{lei.sun@utoronto.ca} \\
}

\begin{document}
\maketitle

\begin{abstract}
\textbf{Motivation:} Accurate power and sample size estimation is crucial to the design and analysis of genetic association studies. When analyzing a binary trait via logistic regression, important covariates such as age and sex are typically included in the model. However, their effects are rarely properly considered in power or sample size computation during study planning. Unlike when analyzing a continuous trait, the power of association testing between a binary trait and a genetic variant depends, explicitly, on covariate effects, even under the assumption of gene-environment independence. Earlier work recognizes this hidden factor but implemented methods are not flexible. \\
\textbf{Method:} We thus propose and implement a generalized method for estimating power and sample size for (discovery or replication) association studies of binary traits that a) accommodates different types of non-genetic covariates E, b) deals with different types of G-E relationships, and c) is computationally efficient. \\
\textbf{Results:} Extensive simulation studies show that the proposed method is accurate and computationally efficient for both prospective and retrospective sampling designs with various covariate structures. A proof-of-principle application focused on the understudied African sample in the UK Biobank data. Results show that, in contrast to
studying the continuous blood pressure trait, when analyzing the binary hypertension trait ignoring covariate effects of age and sex leads to {overestimated power} and {underestimated replication sample size}. \\
\textbf{Availability:} The simulated datasets can be found on the \href{https://github.com/AgueroZZ/Hidden_Factor_Code}{online web-page} of this manuscript, and the UK Biobank application data can be accessed at https://www.ukbiobank.ac.uk. The R package {\tt SPCompute} that implements the proposed method is available at \href{https://cran.r-project.org/web/packages/SPCompute/}{CRAN}. 
\\
\end{abstract}


\section{Introduction}

Accurate power and sample size estimation is crucial to the design of many scientific studies, including the ubiquitous genome-wide association studies (GWAS) of complex and heritable human diseases and traits \citep{hong2012sample}. It is well known that replication studies with underestimated sample sizes can result in false negatives, missing single nucleotide polymorphisms (SNPs; $G$'s) that are truly associated with the phenotype of interest ($Y$) \citep{patil2016should}. Additionally, recent work \citep{MTAG} has shown that failure to correctly estimate power can also result in increased false positives in pleiotropy studies, where different traits are jointly analyzed and their GWAS summary statistics are aggregated. 

The power and sample size calculation for a continuous trait is well established, as the phenotype-genotype association analysis is through the ordinary linear regression, regressing $Y$ on $G$ and important non-genetic covariates $E$'s. It is then straightforward to show that the power of the corresponding genetic association test only depends on the effect size and minor allele frequency (MAF) of the SNP, sample size, and the unexplained phenotypic variance \citep{korte2013advantages}. That is, when analyzing a continuous trait, the sample size for a replication study with sufficient power is determined by the proportion of phenotypic variance explained by genetic variants, which is also called narrow-sense heritability \citep{yang2017concepts, j2017assessing}.

In contrast, the power calculation for a binary disease outcome requires additional considerations, as the association analysis typically uses the logistic or probit regression model \citep{robinson1991some,sjolander2013ignoring}. Most heritability estimation methods were rigorously developed for continuous traits only \citep{weissbrod2018estimating,yang2010common}, and their applications to binary traits have been questioned \citep{golan2014measuring}. At the same time, when analyzing a binary outcome $Y$, power of analyzing a SNP $G$ is affected, explicitly, by the effect size of a non-genetic covariate $E$, even if $E$ is independent of $G$ and/or there is no $G$x$E$ interaction effect \citep{robinson1991some, pirinen2012including}. Therefore, accurate power and sample estimation for a binary trait-genetic association analysis must explicitly consider the presence of non-genetic covariates.

There have been several attempts in the literature to consider the general problem of power and sample size computation for logistic regression. \cite{whittemore1981sample} derived an approximation method, assuming that the disease prevalence is small and the covariates have a joint distribution of multivariate exponential. The approach of \cite{whittemore1981sample} was similarly considered by \cite{hsieh1989sample}, \cite{hsieh1998simple}, and \cite{ novikov2010modified}. Based on the asymptotic power approximation of the score or likelihood ratio test under local alternatives, \cite{self1992power} and \cite{self1988power} proposed an alternative approach that accommodates several categorical covariates with finite configurations, which was then extended by \cite{shieh2000power} to allow for one categorical covariate with infinite configurations.

For genetic association studies, {\tt Quanto} is the most commonly used software in practice, implementing the method of \cite{gauderman2002sample2,gauderman2002sample}. The method uses the expected value of a likelihood ratio test (LRT) statistic and accommodates both continuous and categorical $E$'s for power analysis of $G$x$E$ interaction. However, the approach of \cite{gauderman2002sample} implicitly assumes that $G$ and $E$ are independent of each other, which may not hold in practice for complex diseases \citep{plomin1977genotype,scarr1983people,knafo2013gene, zhu2018causal, namjou2019gwas}. Further, the implemented software {\tt Quanto} does not accommodate the presence of $E$ unless the power computation is for $G$x$E$ interaction analysis. That is, $E$ cannot be included when the power analysis is for the main effect of $G$. 

\cite{demidenko2007sample}, on the other hand, advocated the use of the Wald test to do the power and sample size computation for logistic regression, and proposed a method that allows $E$ and $G$ to be dependent through a second-stage logistic regression model. However, the implemented web-tool \citep{demidenko2008sample} only allows for one binary covariate, as otherwise the computation does not admit a closed-form expression.

\cite{lyles2007practical} proposed a different approach to power computation for generalized linear models, based on the use of an \textit{expanded representative} dataset.  
The idea of expanded representative dataset provides accurate approximation with good computational efficiency when sample size is small to medium, but the computation becomes cumbersome when the sample size is large, while large sample size (and small genetic effect size) is a feature of many GWAS. 

In this paper, we propose and implement a generalized method for estimating power and sample size for genetic association studies of binary traits that a) takes into account different types of non-genetic covariates $E$, b) allows for different types of $G$-$E$ relationship, and c) has good computational efficiency for large-scale studies. 
The utility of the proposed method is illustrated and compared with the existing methods through extensive simulation studies and an application study of the UK Biobank data \citep{sudlow2015uk, bycroft2017genome}. The proposed method has been implemented as a R package, {\tt SPCompute}, available at \href{https://cran.r-project.org/web/packages/SPCompute/}{CRAN}. 

\section{Preliminary}

\subsection{Models}

For simplicity of the notation, we assume without the loss of generality that there is only one non-genetic covariate $E$; the method implementation and application allows for multiple $E$'s. 
To study the relationship between a trait $Y$ and a SNP $G$ of interest, conditional on the non-genetic covariate $E$, we consider the following generalized linear model (glm; \cite{mccullagh2019generalized}), 
\begin{equation} \label{eq:model1}
g(\E(Y|X)) = g(\mu) = \beta_0 + \beta_G G + \beta_E E = \eta,    
\end{equation}
where $g(\cdot)$ is a link function, connecting the linear predictor $\eta$ with the mean function of $Y$. This glm model accommodates the analyses of both continuous and binary traits. Here we focus on binary traits, for which the logistic regression is the most commonly used model with $g(\mu) = \log(\frac{\mu}{1-\mu})$.

Let ${X}$ be the design matrix, which has rows $\{(1,G_i,E_i)\}_{i=1}^n$, where $n$ is the sample size.  To ease notation, we also use $X$ to denote the observed data, and we use $\beta=(\beta_0,\beta_G,\beta_E)^T$ to denote the vector of either all regression parameters or their true values. The linear predictor $\eta$ is then expressed as
$$\eta = X \beta := \beta_0 + \beta_G G + \beta_E E.$$ 
Following the convention in genetic association studies, the SNP genotypes, $aa$, $Aa$ and $AA$, are assumed to follow the Hardy--Weinberg equilibrium (HWE; \cite{mayo2008century}), where $A$ is the minor allele with MAF of $p$. Also by convention, $G$ is coded additively, tracking the number of allele $A$. Thus, $\p(G = 0) = (1-p)^2$, $\p(G=1) = 2p(1-p)$ and $\p(G=2) = p^2$.

\subsection{The Wald test}\label{section:wald}

To test the association between SNP $G$ and trait $Y$, i.e.\ $H_0: \beta_G = 0$ vs.\ $H_1: \beta_G \neq 0$, one can consider different tests such as the LRT, Score or Wald tests. These tests have similar asymptotic behaviors under the null hypothesis, and they are locally equivalent 
\citep{rao1973linear, serfling2009approximation}. However, as noted by \cite{demidenko2007sample}, these three likelihood-based tests differ globally. 
As Wald tests are routinely used in GWAS, following the argument of \cite{demidenko2007sample}, we carry out our power and sample size computation based on a Wald test.

The Wald test statistic in our setting is expressed as 
$$T = \frac{{\hat{\beta}_G}^2}{{I_{{X}}^{-1}(\hat{\beta})_{[2,2]}}},$$ 
where $I_{{X}}^{-1}(\hat{\beta})_{[2,2]}$ denotes the second diagonal element of the matrix $I_{{X}}^{-1}(\hat{\beta})$, and $\hat{\beta}$ is the maximum likelihood estimate (MLE) of $\beta$. $I_{{X}}(\beta)$ is the observed or conditional Fisher information matrix, conditional on the observed $X$,  defined as
$$I_{{X}}(\beta) = {X}^TW(\beta) {X},$$ 
where $W(\beta)$ is a $n\times n$ diagonal matrix, with the $i^{th}$ element as 
$$w_i=(\frac{\partial u_i}{\partial\eta_i})^2/\Var(Y_i|X_i).$$ 
Under the null hypothesis, $T$ is asymptotically $\chi^2_1$ distributed. 

Using the above expression of $w_i$, it is easy to see that when analyzing a continuous trait with residual variance $\sigma^2$ via linear regression (i.e.\ using the identity link function), $\frac{\partial u_i}{\partial\eta_i}=1$ and  $w_i = 1/\sigma^2$, which means $I_{{X}}(\beta)$ depends on $\sigma^2$ but not on any regression coefficients explicitly. In contrast, when analyzing a binary trait using logistic regression, 
$$w_i = \frac{\exp(-\eta_i)}{(1+\exp(-\eta_i))^2},$$ 
which is a function of both $\beta_G$ and $\beta_E$. Thus, the size of the non-genetic covariate effect $\beta_E$ explicitly influences the Fisher information matrix, hence the power analysis of  $\beta_G$. 

\subsection{The hidden factor in power and sample size computation}

Assume the significance level of the test is $\alpha$, and the sample size is large enough so that the asymptotic distribution of the Wald test statistic can be used. 
Let 
$${V}_{G,{X}}:= I_{{X}}^{-1}(\beta)_{[2,2]}$$ 
be the variance of $\hat{\beta}_G$,  the power of the Wald test can be computed as 
$$\Phi(-Z_{1-\alpha/2} + \frac{\beta_G}{\sqrt{V_{G,{X}}}}) + \Phi(-Z_{1-\alpha/2} - \frac{\beta_G}{\sqrt{V_{G,{X}}}}),$$ 
where $Z_{1-\alpha/2}$ denotes the $1-\alpha/2$ quantile of the standard normal distribution. Worth reemphasizing is the fact that ${V}_{G,{X}}$, thus the power of logistic regression, explicitly depends on both $\beta_G$ and $\beta_E$,  as discussed in Section \ref{section:wald} above.

The above power computation is for \textit{conditional} power, conditional on the observed ${X}$. However, for sample size determination for a successful replication study, the corresponding power analysis is performed prior to observing any data. In that case, the power is referred to as the \textit{unconditional} power \citep{lyles2007practical}. 

To compute the unconditional power, naturally we replace the conditional Fisher information matrix $I_{{X}}(\beta)$ above with its unconditional version $I_n(\beta)$. Let $I_{\bd x} (\beta)$ be the unconditional Fisher information for a single observation $\bd{x} = (1,G,E)'$. For the logistic regression considered, 
{\small 
\begin{equation} \label{eq:fisher}
\begin{aligned}
 & I_{1}(\beta) = \E_{\mathbb{F}_X} [w\bd{x}^T\bd{x}] \\
                 &=  \E_{\mathbb{F}_X} \bigg[ \frac{\exp(-(\beta_0 + \beta_G G + \beta_E E)}{(1+\exp(-(\beta_0 + \beta_G G + \beta_E E)))^2} \begin{bmatrix}
1  & G & E \\
G & G^2 & GE \\
E & GE & E^2 
\end{bmatrix}
\bigg],
\end{aligned}
\end{equation}
}
where the expectation is taken over, $\mathbb{F}_X$, the distribution of the covariate space $X$ (i.e.\ both $G$ and $E$).

Once $I_{1} (\beta)$ has been computed for a given $\mathbb{F}_X$, the unconditional Fisher information matrix for a random sample of size $n$ is
$$I_n(\beta) := n I_{1} (\beta).$$ 
The unconditional power is then
\begin{equation}\label{powerformula}
\begin{aligned}
\Phi(-Z_{1-\alpha/2} + \frac{\beta_G}{\sqrt{V_{G,n}}}) + \Phi(-Z_{1-\alpha/2} - \frac{\beta_G}{\sqrt{V_{G,n}}}),
\end{aligned}
\end{equation}
where 
$$V_{G,n}: = I_n^{-1}(\beta)_{[2,2]}$$ 
is based on the unconditional Fisher information matrix, $I_n(\beta)$. 

To plan a successful replication study at the $\alpha$ level, the sample size $n$ required to achieve a desirable power can be computed by simply inverting the power function, which is monotonic with respective to $n$. Although the sample size computation is for a specific genetic effect $\beta_G$, it is clear that, similar to the conditional Fisher information, the unconditional Fisher information in (\ref{eq:fisher}), therefore $V_{G,n}$ in (\ref{powerformula}), also depends $\beta_E$. 
Thus, this hidden factor must be explicitly accounted for when performing sample size  calculation for a binary trait.

\section{Methods}

\subsection{Designing the covariate space}\label{cov-space}
To compute the unconditional Fisher information matrix $I_n(\beta)$, one needs to compute the moments and covariance of a random sample pair $(G_i,E_i)$ from the corresponding covariate space $\mathbb{F}_X$. 
An appropriately designed covariate space $\mathbb{F}_X$ should be flexible enough to accommodate potential complex dependence structure between $G$ and $E$, while conceptually simple enough so that that practitioners can make use of their domain knowledge.

In the work of \cite{gauderman2002sample}, the author implicitly assumed independence between $G$ and $E$, requiring only the marginal distributions of $G$ and $E$. 
Although this makes the method easy-to-implement, the assumption may not hold in practice \citep{plomin1977genotype,scarr1983people,knafo2013gene, zhu2018causal, namjou2019gwas}. Furthermore, the implemented software {\tt Quanto} only allows users to specify $E$ when the target analysis is the $G\times E$ interaction effect, not the main effect of $G$. 

The work of \cite{demidenko2007sample}, on the other hands, allows $\mathbb{F}_X$ to accommodate dependence between a binary $G$ and a binary $E$, by introducing a second-stage logistic regression, 
\begin{equation}\label{ori-second-stage}
\begin{aligned}
\log\bigg(\frac{\p(G=1|E)}{\p(G=0|E)}\bigg) = \gamma_0^* + \gamma_E^* E,
\end{aligned}
\end{equation}
where $\gamma_0^*$ is determined by user-specified marginal probabilities of $G$ and $E$. As a result, users will only need to additionally input the knowledge about $\gamma_E^*$ to fully specify $\mathbb{F}_X$. However, the method of \cite{demidenko2007sample} is designed for a binary $G$ (hence the typical GWAS additive coding of $G$ not applicable) and a binary  $E$, and its generalization to different types of $G$ and $E$ is nontrivial. 

Here, we utilize the idea of a second-stage regression of \cite{demidenko2007sample} but extend it to a more general setting. Instead of treating $E$ as a covariate in the second-stage regression, we consider it to be the response variable such that,
\begin{equation}\label{new-second-stage}
\begin{aligned}
g_2(\E(E|G)) = \gamma_0 + \gamma_G G,
\end{aligned}
\end{equation}
where $g_2$ is the link function, being identity when $E$ is continuous and logit when $E$ is binary. 

Compared with the second-stage regression model in (\ref{ori-second-stage}), the proposed method can accommodate different types of $E$ in a unified framework. 
When $E$ is continuous, the regression model also requires $\Var(E|G)$ in order to be fully specified. The value of $\Var(E|G)$ can be computed based on user-provided information such as $\mu_E, \sigma_E$ and $p$. 

{We note that the proposed method can account for multiple $E$'s by specifying the corresponding $g_2$ function for each $E$ considered. Later, we will demonstrate the utility of our approach in a UK Biobank data application of hypertension, for which both age and sex are important covariates to consider for power and sample size computation. In the rest of this section, we assume there is only one covariate $E$ for clarity of the presentation.}

\subsection{Proposed method 1: Semi-Simulation (P1.SS)}\label{m1}

The estimation of the unconditional power heavily depends on the computation of $I_n(\beta) := n I_{1} (\beta)$. Unfortunately, unless in some special cases such as when both $G$ and $E$ are binary, $I_{1} (\beta)$ in (\ref{eq:fisher}) does not have a closed-form expression for a general $\mathbb{F}_X$ \citep{demidenko2007sample}. Thus, to estimate $I_{n} (\beta)$, we propose to use a sample estimate.

Specifically, for a large integer $B$, we simulate independent observations $\{G_i, E_i\}_{i=1}^B$ from the covariate space $\mathbb{F}_X$, and for each $\bd x_i = (1, G_i, E_i)'$ we compute the corresponding conditional Fisher information matrix,
$$I_{\bd x_i}(\beta) = {\bd x_i}^Tw_i(\beta) {\bd x_i},$$ 
where $$w_i(\beta) = \frac{\exp(-(\beta_0 + \beta_G G_i + \beta_E E_i)}{(1+\exp(-(\beta_0 + \beta_G G_i + \beta_E E_i)))^2} .$$  By a simple application of the law of large number, the sample estimate, 
\begin{equation}\label{eq:6}
\begin{aligned}
\tilde{I}_n(\beta) := \frac{\sum_{i=1}^{B} n I_{\bd x_i}(\beta)}{B},
\end{aligned}
\end{equation}
converges almost surely to the true expected matrix $I_n(\beta)$ as $B$ grows. 

As we will illustrate later in the simulation studies, $\tilde{I}_n(\beta)$ exhibits little variation for large $B$ (e.g.\ >10,000). Furthermore, the proposed \textit{semi-simulation} method is scalable, as for each $I_{\bd x_i}(\beta)$ we only compute the observed Fisher information matrix for one single observation. Thus, the computational load depends on $B$ but is independent of the target sample size $n$. 
Once $I_n(\beta)$ is replaced by  $\tilde{I}_n(\beta)$, the power computation can proceed using equation (\ref{powerformula}), and sample size estimation by inverting the power function.

\subsection{Proposed method 2: Representative Dataset (P2.RD)}\label{m2}

An alternative method that does not rely on plugging in the sample estimate of $I_n(\beta)$ is through the use of a \textit{representative} dataset, an idea that was originally suggested by \cite{o1986using} and later extended by \cite{lyles2007practical}. 

In our setting, given a sample size $n$, assume there exists a representative covariate sample $ \{\boldsymbol{x}_i\}_i^n=\{(1, G_i,E_i)'\}_{i=1}^n$ from the covariate space $\mathbb{F}_X$, which we define later. We then expand $ \{\boldsymbol{x}_i\}_i^n$ to consider both possible outcomes of the binary trait, so that each observation $\boldsymbol{x}_i$ splits into $\{\boldsymbol{x}_i, y_i = 0\}$ and $\{\boldsymbol{x}_i, y_i = 1\}$. Additionally, each expanded observation is given a weight, so that $\delta_i^0+\delta_i^1=1$, where
\begin{equation} \label{eq:weight}
     \delta_i^l = \p(y_i = l|\boldsymbol{x}_i) 
                = \frac{\exp(\beta_0 + \beta_G G_i + \beta_E E_i)^l}{1+\exp(\beta_0 + \beta_G G_i + \beta_E E_i)},
\end{equation}
for $l =$ 0 and 1.

Thus, the original representative dataset $\{\boldsymbol{x}_i\}_{i=1}^{n}$ is now expanded into the following representative dataset,
\begin{equation} \label{eq:rd}
\begin{aligned}
RD:= \begin{Bmatrix}
\boldsymbol{x}_i, & y_i = 0, & \delta_i^0 \\
\boldsymbol{x}_i, & y_i = 1, & \delta_i^1 \\
\end{Bmatrix}_{i=1}^{n},
\end{aligned}
\end{equation}
which has $2n$ (weighted) observations. Standard MLE of $V_{G,n}$, derived from the corresponding weighted log-likelihood, yields $\hat{V}_{G,n}$, which can be directly plugged into equation (\ref{powerformula}) to complete the power computation \citep{lyles2007practical}.

It remains to be discussed what is a representative $\{\boldsymbol{x}_i\}_{i=1}^n$ and how the expanded representative dataset $RD$ can be obtained in our study setting. In the case of conditional power analysis where covariates are already observed, the observed $\{\boldsymbol{x}_i\}_{i=1}^n$ can be directly used in (\ref{eq:weight}) to establish the representative dataset of (\ref{eq:rd}). 

For the unconditional power analysis, $\{\boldsymbol{x}_i\}_{i=1}^n$ can be obtained by using user-provided $\mathbb{F}_X$. \cite{lyles2007practical} provided examples on how to define the notion of representative dataset for different types of $\mathbb{F}_X$. We follow the procedures of \cite{lyles2007practical} for the types of $\mathbb{F}_X$ considered in Section \ref{cov-space}.

When $E$ is binary and the link function in  (\ref{new-second-stage}) is logistic, we can compute the expected counts for category $\{(G = i, E = j)\}$, $i=0, 1$, and $2$, and $j=0$ and $1$, as
$$n_{i,j} = n\p(G = i, E = j),$$ 
using the available information such as MAF and the inheritance mode,  with appropriate rounding to ensure that $n_{i,j}$'s are integers and sum to $n$. 

When $E$ is continuous and the link function is identity with $\Var(E|G) = \sigma_E^2$, we first categorize the dataset based on $G$ such that $n_i = n \p(G = i)$, $i=0, 1$, and 2. Then for each of the $j=1,\ldots, n_i$ observations of $\{G_j = i\}_{j=1}^{n_i}$, 
$$E_j = \gamma_0 + \gamma_G i + \sigma_E \Phi^{-1}[(j-0.375)/(n_i + 0.25)],$$
where $\Phi^{-1}$ is the inverse of the cumulative distribution function of the standard normal. 

\section{Simulation Studies}\label{sim}

\subsection{Overview of the simulation design}\label{sim1}

We compared the power and sample size computed using the proposed  P1.SS and P2.RD methods (implemented as the R package {\tt SPCompute} available at CRAN) with those computed using {\tt Quanto} of \cite{gauderman2002sample} (version 1.2.4 downloaded from http://hydra.usc.edu/GxE), and the method of \cite{demidenko2007sample} using its web-platform (dartmouth.edu/$\sim$eugened/power-samplesize.php). 

We considered three different scenarios for $\mathbb{F}_X$, including no covariate $E$ (as {\tt Quanto} does not allow for $E$), $E$ being binary (as the method of \cite{demidenko2007sample} only allows for binary $E$), and $E$ being continuous. {Although we only considered one $E$ in the simulation studies for method comparison, our implemented {\tt SPCompute} R package allows for multiple $E$'s, as demonstrated in our UK Biobank application study in Section \ref{real}.} Finally, for the simulation studies we also  considered three study designs, where S1 is case-control retrospective  ({\tt Quanto} only allows for the case-control study design), while S2 and S3 are  prospective to reflect the design of the emerging biobank-sized data such as the UK Biobank data used in our application. 

The accuracy of each method was assessed by comparing the computed power (and sample size) with the empirical values obtained through 1,000 independent replications. Given the large number of replications, the empirical values were treated as the oracle values and used to benchmark. We calculated the average and the maximum of the absolute error (AE) of each computed power (and sample size) as compared with the oracle values. The more accurate method is expected to have smaller mean AE and max AE. By convention, for replication sample size computation, the desirable power was set to be 80\% at the significance level of $0.05$, and for consistency, the same significance level was used for power computation.

\subsubsection{Scenario 1: No covariate $E$ with a case-control retrospective study design} \label{sec:S1}

The choice of no covariate effect was to accommodate the implementation of {\tt Quanto} of \cite{gauderman2002sample}. Without loss of generality, the disease prevalence was assumed to be 20\%, and the observations were obtained independently with a retrospective sampling design and the standard case-to-control ratio of 1-to-1. 

The associated SNP has a MAF of $0.1$, with a dominant  effect $\beta_G$ ranging from $\log(1.1)$ to $\log(2.5)$. That is, the odds ratio (OR) of the associated SNP ranged from 1.1 to 2.5. The choice of a dominant genetic effect was to accommodate the implementation of \cite{demidenko2007sample} method, which only allows for a binary $G$.  Finally, we used $\beta_0 = -2$, though we note that the intercept parameter does not affect the power of a  case-control study. 

\subsubsection{Scenario 2: Binary $E$ with a prospective study design} \label{sec:S2}

Similar to S1 above, in the second scenario the disease prevalence is also 20\%, and the associated SNP with MAF of $0.1$ has a dominant effect $\beta_G$ ranging from $\log(1.1)$ to $\log(2.5)$. However, the observations were obtained independently with a prospective sampling design as in the UK Biobank data. Additionally, the non-genetic covariate $E$ has a population exposure rate of $\p(E=1) = 0.3$ with effect $\beta_E = \log(2.5)$. Finally, 
$\gamma_G=\log(0.2)$, quantifying the dependency between $G$ and $E$ as defined in (\ref{new-second-stage}). 

To implement {\tt Quanto} of \cite{gauderman2002sample}, which only allows for case-control study design, we used a case-to-control ratio of 1-to-4 to approximate the result for a disease with prevalence of 20\%. Additionally, when the power analysis is about $G$ main effect (as opposed to $G\times E$ interaction effect), {\tt Quanto} does not consider the presence of $E$. Thus, we only input the information about $G$ in the implementation of {\tt Quanto}. The method of \cite{demidenko2007sample} accommodates the presence of one binary covariate $E$ for the power (and sample size) computation; $G$ must be binary, hence the dominant genetic model was assumed for method implementation.



\subsubsection{Scenario 3: Continuous $E$ with a prospective study design} \label{sec:S3}

For this last scenario, without loss of generality, the covariate $E$ was assumed to follow the standard normal distribution conditional on $G$. The dependency between $G$ and $E$ was set to $\gamma_G=\log(0.5)$. All other model specifications are the same as in S2 above, including the disease prevalence (20\%), MAF (0.1), the genetic effect (ranging from $\log(1.1)$ to $\log(2.5)$), and the non-genetic covariate effect ($\log(2.5)$.

As in the previous scenario, we ignored the information on $E$ for the implementation of {\tt Quanto}. For the method of \cite{demidenko2007sample}, which only allows for a binary $E$, we considered two approaches. We first omitted the continuous covariate $E$ (corresponding results$^*$), and we then dichotomized $E$ by defining $\tilde{E} := \mathbb{I}(E>0)$. This corresponds to creating two misspecified models, 
\begin{equation}\label{miss1}
\begin{aligned}
\log \bigg(\frac{\p(Y=1|G,\tilde{E})}{\p(Y=0|G,\tilde{E})}\bigg) &= \beta_0 + \beta_G G + \tilde{\beta}_E \tilde{E},
\end{aligned}
\end{equation}
and 
\begin{equation}\label{miss2}
\begin{aligned}
\log \bigg(\frac{\p(\tilde{E}=1|G)}{\p(\tilde{E}=0|G)}\bigg) &= \tilde{\gamma}_0 + \tilde{\gamma}_G G.
\end{aligned}
\end{equation}
As the parameter values specified for the true models (\ref{eq:model1}) and (\ref{new-second-stage}) cannot be directly used for the two misspecified models, we used estimated $\tilde{\gamma}_G$ and $\tilde{\beta}_E$. We first simulated a large a number of observations $\{G_i, E_i, Y_i\}_{i}^{3\times 10^5}$ using the true model. We then dichotomized the continuous $E$ to obtain $\tilde{E}$ as specified above. Finally, we regressed $Y$ on $G$ and $\tilde{E}$, and $\tilde{E}$ on $G$ to obtain sample estimates of $\tilde{\beta}_E$ and $\tilde{\gamma}_G$ for the second implementation of the method of \cite{demidenko2007sample}.  



\subsubsection{Methods comparison across the three scenarios} \label{sec:results}

\begin{figure}[p]
\centering
\subfigure[S1:\ Retrospective without covariate]{
  \includegraphics[width=0.46\textwidth]{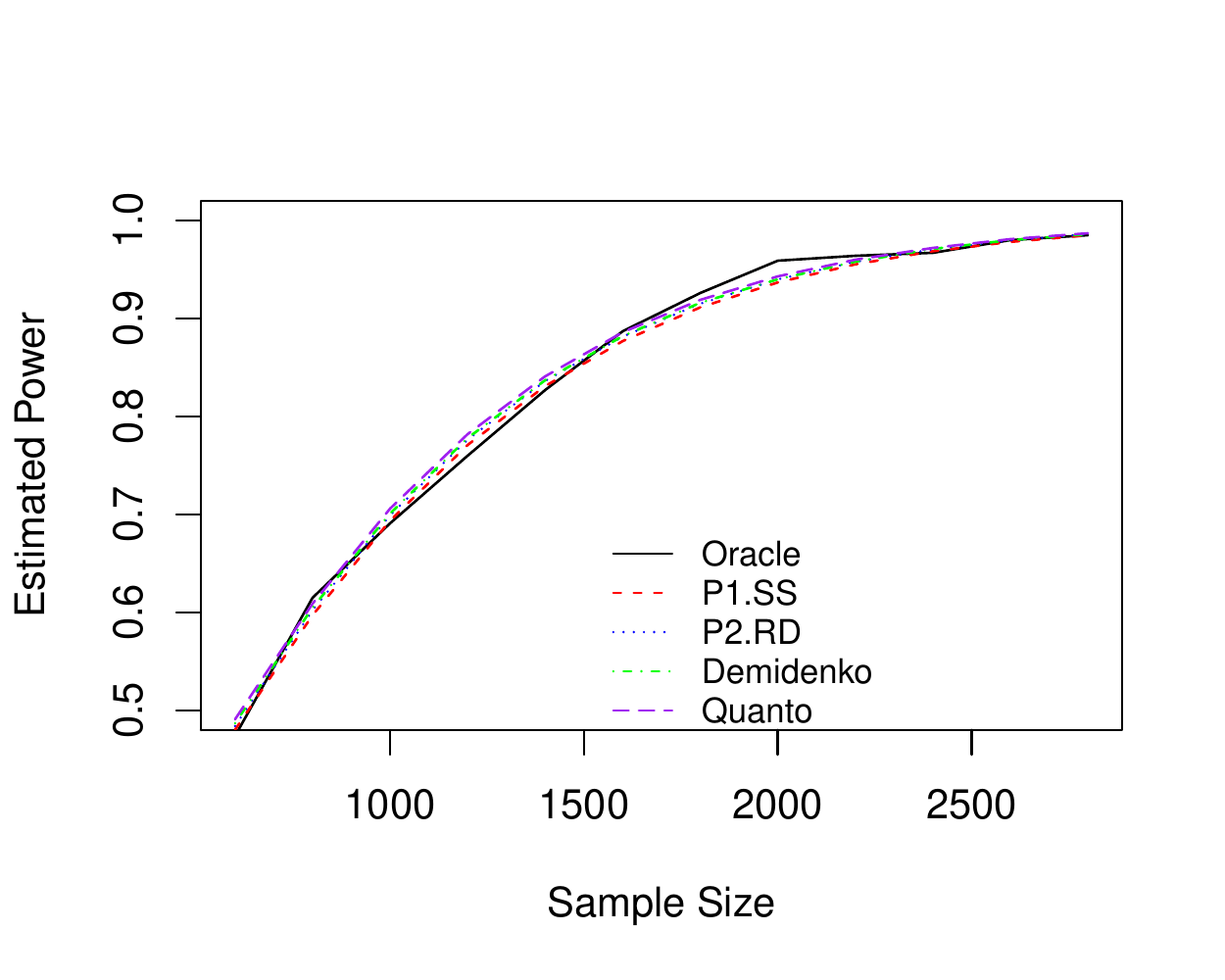}
}
\subfigure[S1:\ Retrospective without covariate]{
	\includegraphics[width=0.46\textwidth]{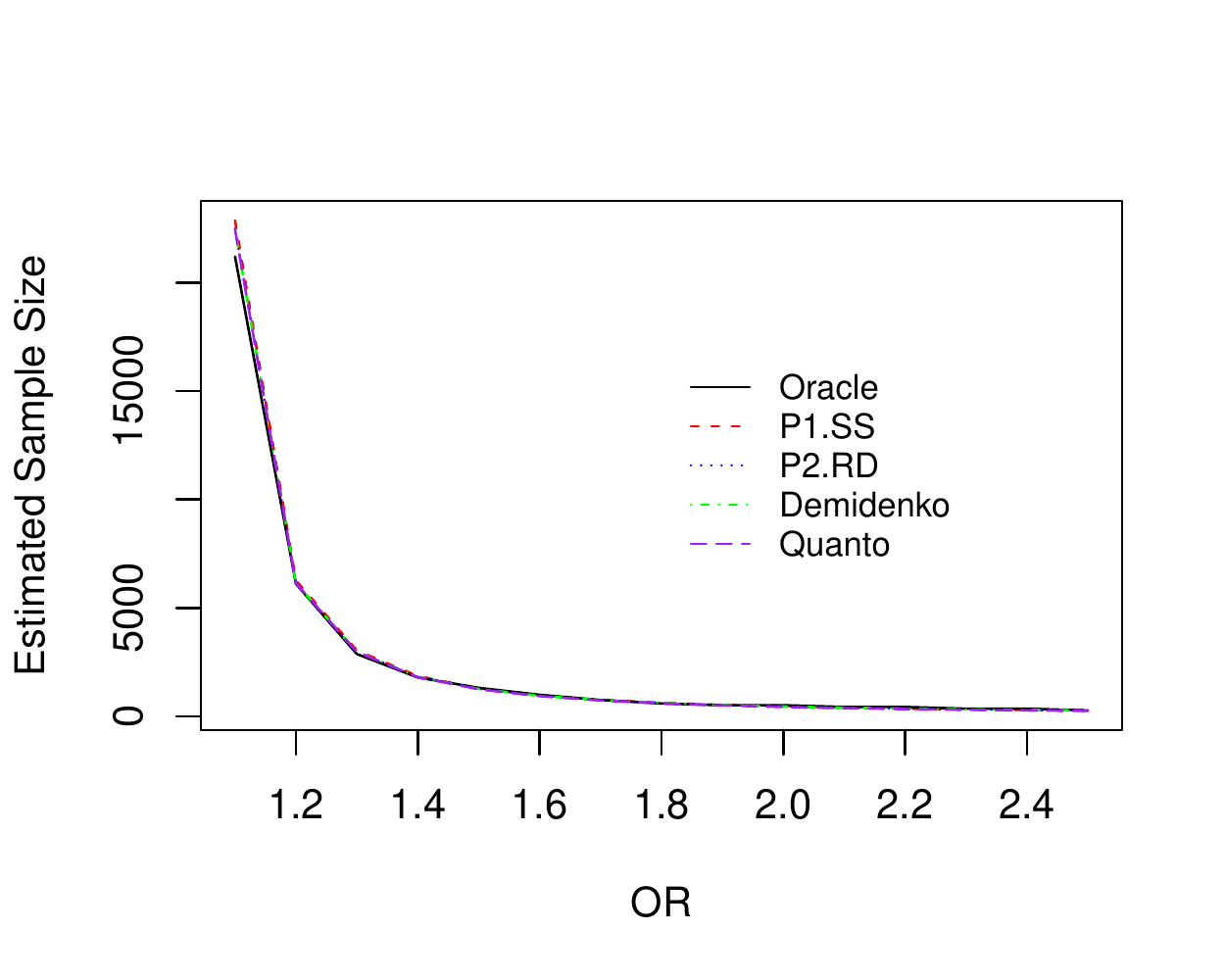}
}
\subfigure[S2:\ Prospective, binary covariate]{
  \includegraphics[width=0.46\textwidth]{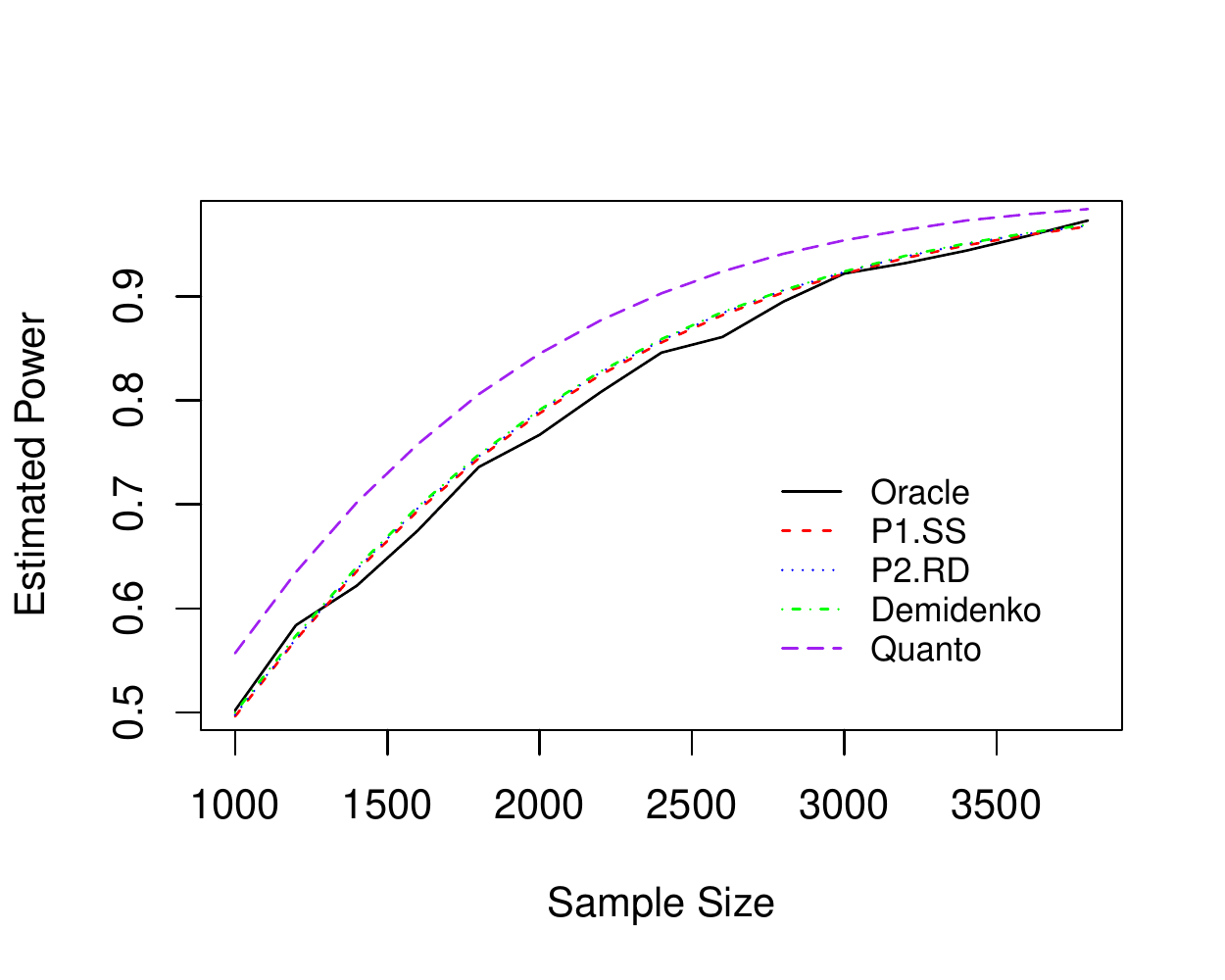}
}
\subfigure[S2:\ Prospective, binary covariate]{
	\includegraphics[width=0.46\textwidth]{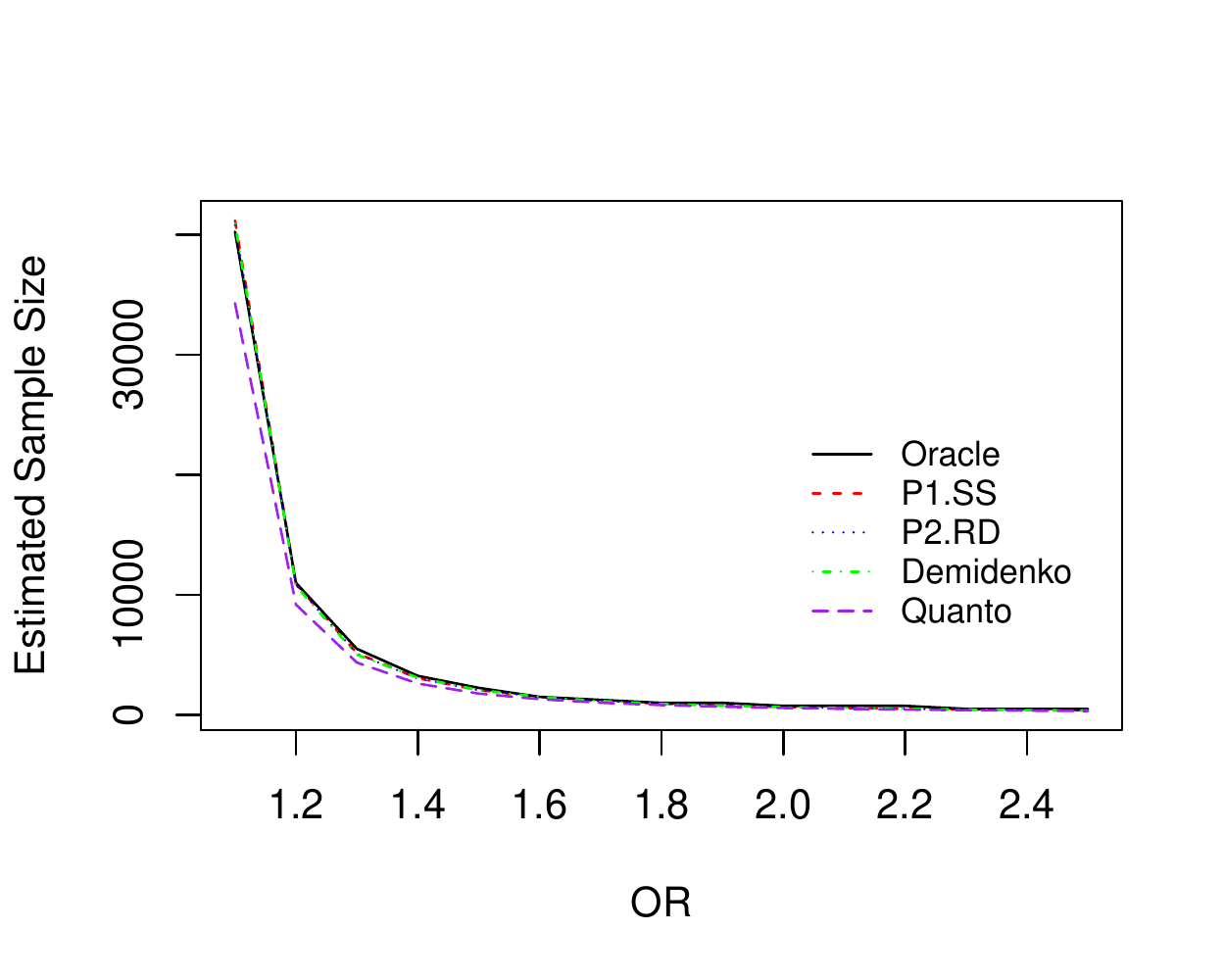}
}
\subfigure[S3:\ Prospective, continuous covariate]{
  \includegraphics[width=0.46\textwidth]{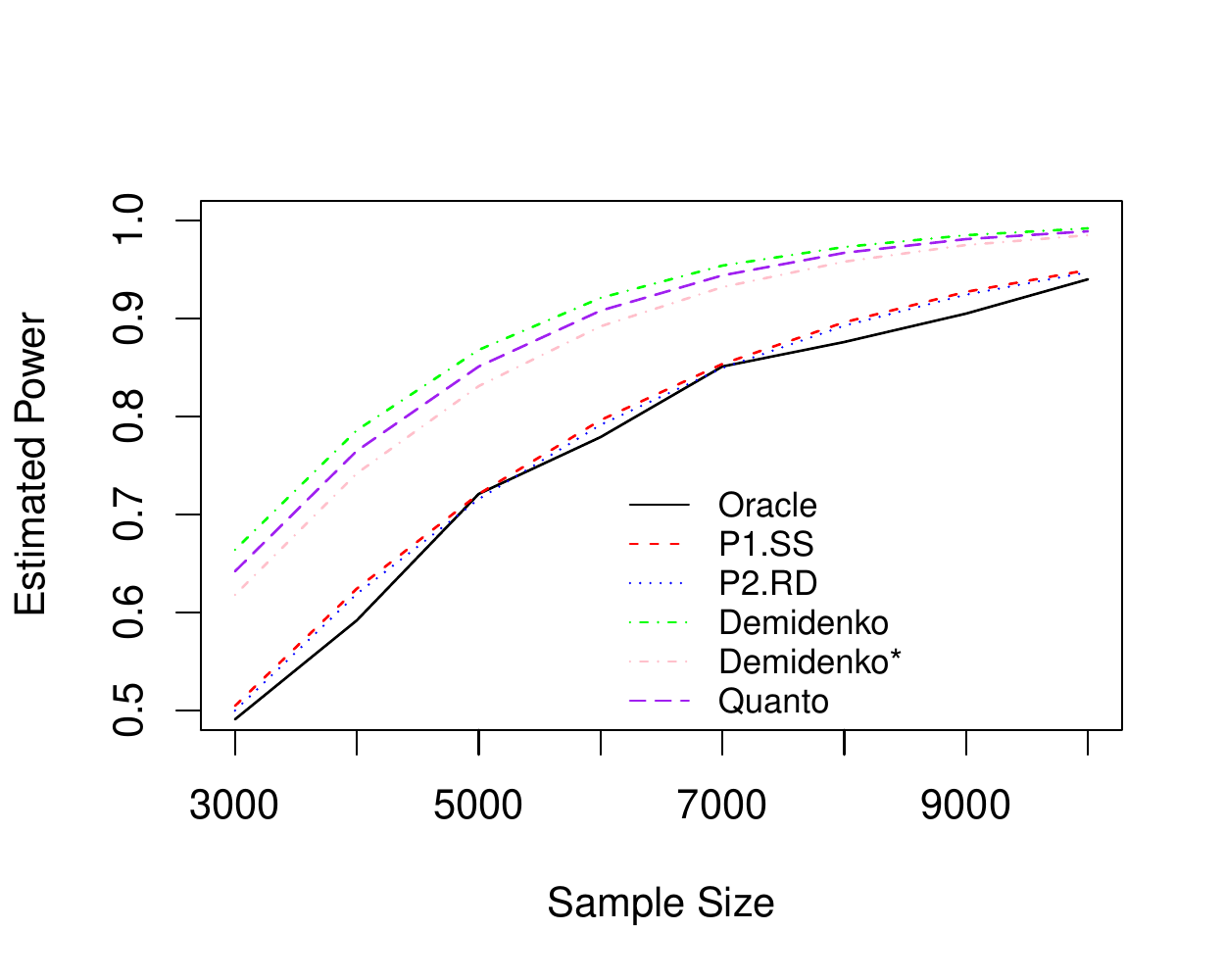}
}
\subfigure[S3:\ Prospective, continuous covariate]{
	\includegraphics[width=0.46\textwidth]{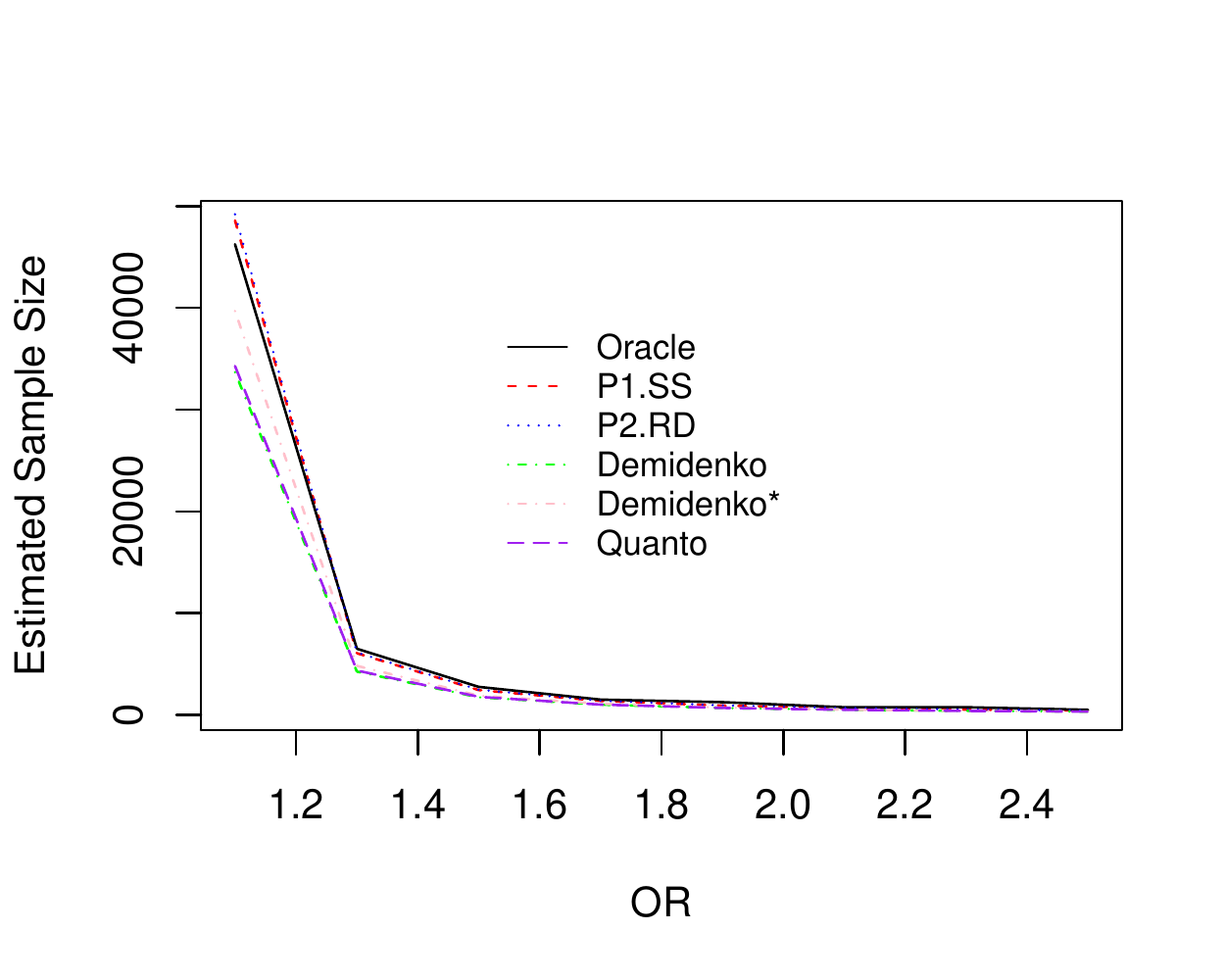}
}
\caption{Simulation results for the three scenarios considered in Section \ref{sim1}. Scenario 1 (S1) is the retrospective case-control sampling design without $E$. Scenario 2 (S2) and Scenario 3 (S3) are the prospective sampling design with, respectively, a binary and continuous covariate $E$. Figures (a), (c) and (e) on the left panel compare the power computation when $\beta_G$ is fixed at log(1.5) (i.e.\ OR of 1.5, for (a-c)) or log(1.3) (for (e)), and Figures (b), (d) and (f) on the right panel compare the sample size computation to achieve power of 80\% at the significance level of 0.05 across different $\exp(\beta_G)$. The red curves are for the `semi-simulation' method (P1.SS in Section \ref{m1}), blue curves for the `representative dataset' method (P2.RD in Section \ref{m2}), purple curves for {\tt Quanto} of \cite{gauderman2002sample}, and green and pink curves for the method of \cite{demidenko2007sample}; in S3, the method of \cite{demidenko2007sample} was implemented by dichotomizing $E$ or without considering $E$). The black cures represent the oracle power and replication sample size estimated empirically.}
\label{fig:Sim1Figures}
\end{figure}

Figure \ref{fig:Sim1Figures} shows the computed power and sample size curves, across the three scenarios, and the empirical results are consistent with our analytical expectation: Ignoring covariate effect, {\it at the (replication) study planning stage},  can lead to {overestimated power} and {underestimated replication sample size} for studying a binary trait. 

In the left panel of Figure \ref{fig:Sim1Figures}, the OR of the associated SNP was fixed at 1.5 in (a,c) and 1.3 in (e), and Figures \ref{fig:Sim1Figures}(a), \ref{fig:Sim1Figures}(c) and \ref{fig:Sim1Figures}(e) show the estimated power for different sample size at the significance level of 0.05. Results clearly show that, if there are no covariates, all methods perform well (Figure \ref{fig:Sim1Figures}(a)). In the presence of a binary covariate, {\tt Quanto} tends to {\it overestimate} the power of an association study, while both Demidenko and the proposed two methods provide power estimates close to the Oracle values (Figure \ref{fig:Sim1Figures}(c)). However, if the influential covariate is continuous (e.g.\ age as in our UK Biobank application for hypertension), only the proposed two methods (P1.SS and P2.DD) perform well (Figure \ref{fig:Sim1Figures}(e)). The superior performances of the proposed two methods are also shown in Table \ref{table:Sim1table}, as P1.SS and P2.RD have smaller average and smaller maximum absolute error, as compared with the true power. 

In the right panel of Figure \ref{fig:Sim1Figures}, Figures \ref{fig:Sim1Figures}(b), \ref{fig:Sim1Figures}(d) and \ref{fig:Sim1Figures}(f) show the estimated sample sizes, necessary to achieve 80\% power at the 0.05 level, to successfully replicate an associated SNP with OR ranges from 1.1 to 2.5. Similar conclusion can be drawn here, as the existing methods tend to {\it underestimate} the necessary sample size for a successful replication study in the presence of influential covariate, while the proposed P1.SS and P2.RD methods are accurate.

\subsection{Choice between the proposed P1.SS and P2.RD methods from the computational perspective}\label{selection}

To select between the two proposed methods, P1.SS and P2.RD as respectively described in Section \ref{m1} and Section \ref{m2}, here we study factors influencing the computational efficiencies of the two methods and make recommendations to practitioners. 

Conceptually, the computational efficiency of the semi-simulation P1.SS method depends on $B$, the number of independent observations drawn from $\mathbb{F}_X$ in order to obtain 
$\tilde{I}_n(\beta)$ in (\ref{eq:6}). As $\tilde{I}_n(\beta)$ is based on $B$ replicates of one-sample $I_{\bd x_i}(\beta)$, the targeted sample size $n$ does not have a direct impact on computational time. In contrast, the computational time of the P2.RD method depends on $n$, as the method first creates a representative dataset of size $n$ from $\mathbb{F}_X$ then expand it to weighted $2n$ observations as in (\ref{eq:rd}). 

\begin{figure}[p]
\centering
\subfigure[Runtime comparison]{
	\includegraphics[width=0.46\textwidth]{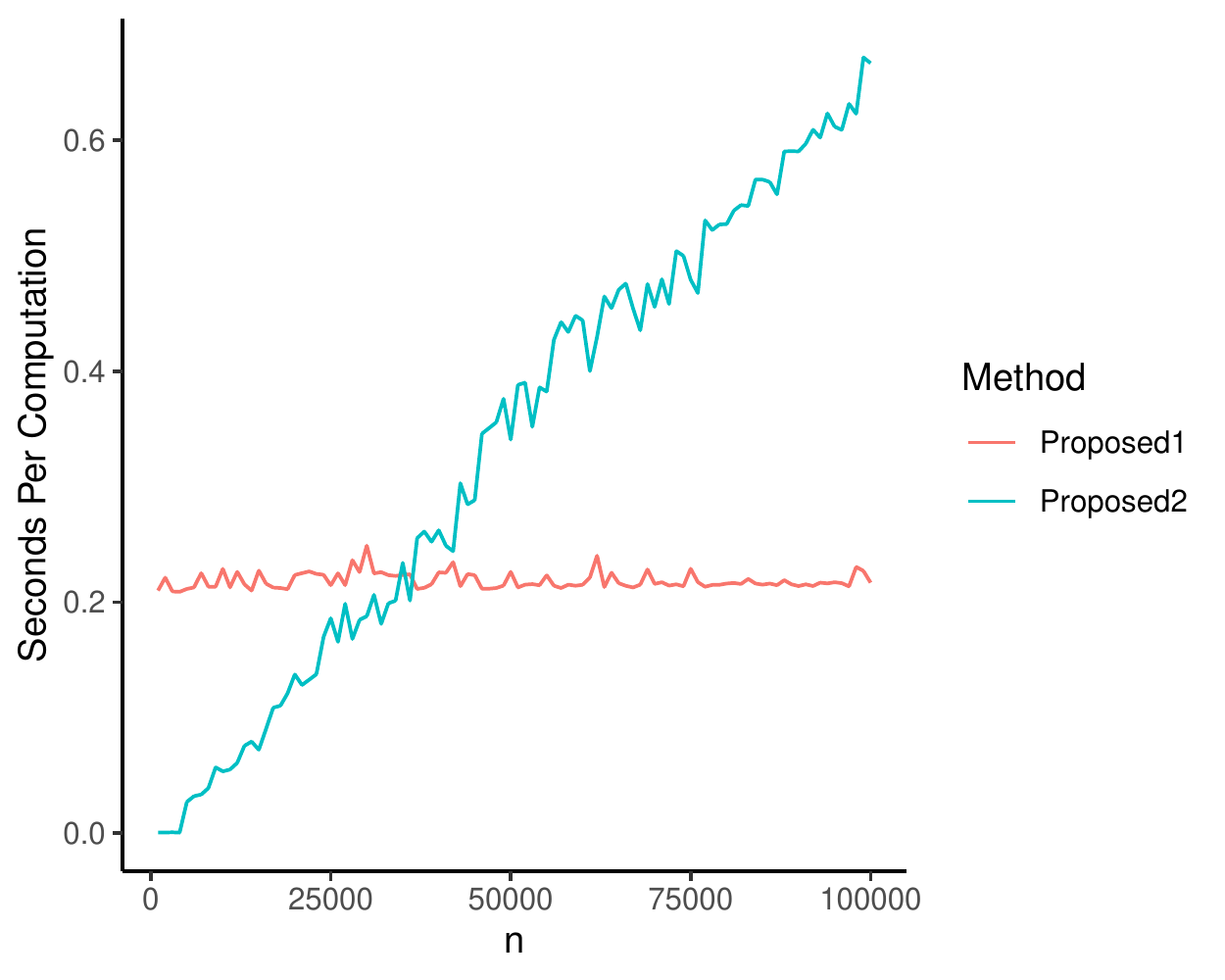}
}
\subfigure[Stability of P1:SS]{
  \includegraphics[width=0.46\textwidth]{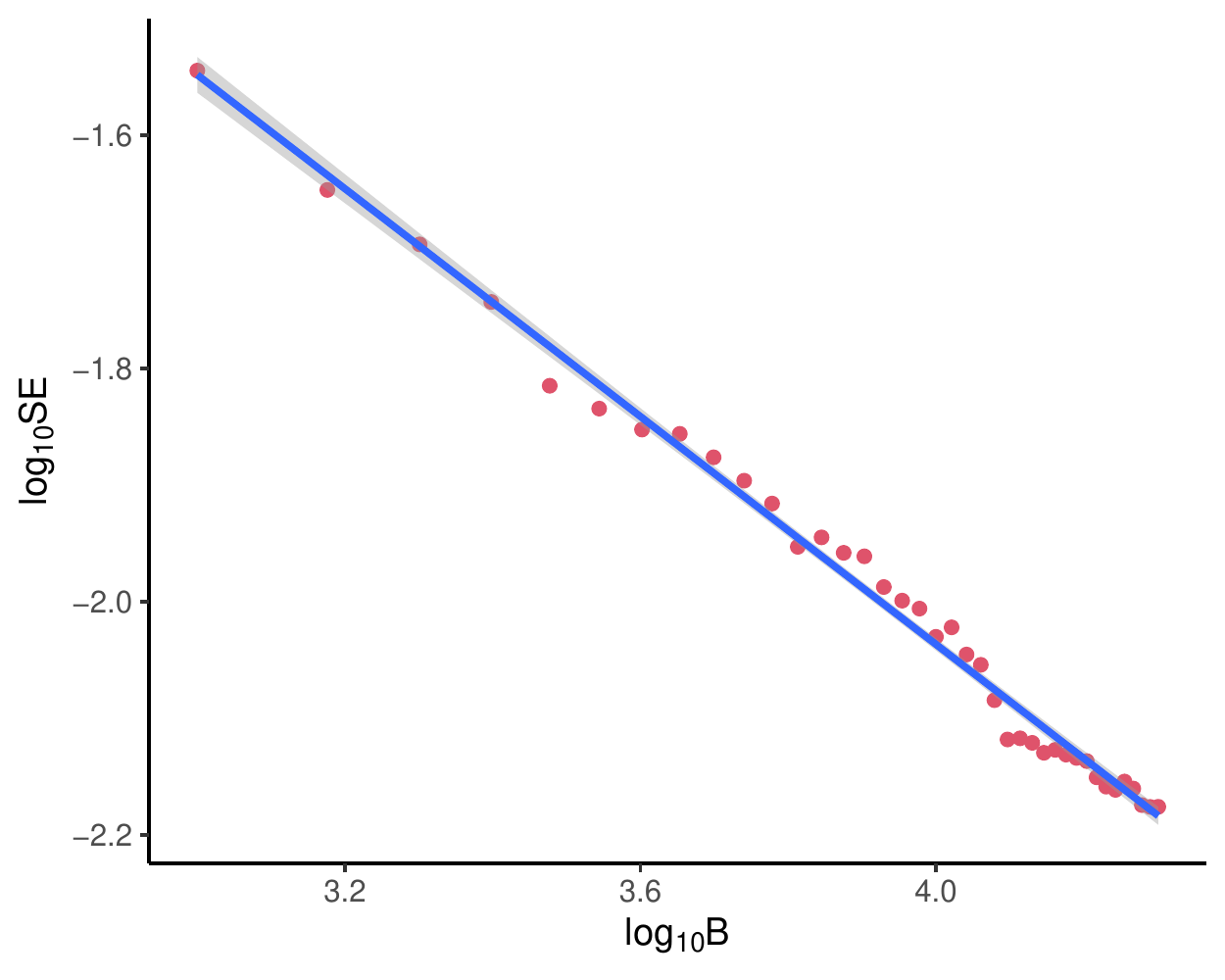}
}
\caption{Figure (a) displays the relationship between the run time in seconds per computation for each method across different sample sizes, using simulation Scenario 2; results for the other two scenarios are characteristically similar. The Proposed 1 is the `semi-simulation' method (P1:SS) in Section \ref{m1}, The proposed 2 is the `representative dataset' method (P2:RD) in Section \ref{m2}. Figure (b) shows that there is a linear relationship between the $\log_{10}$ standard error (SE) of estimated power and the $\log_{10}$ number of replicates (B) used for the proposed `semi-simulation' method 1. }
\label{fig:Sim2Figure}
\end{figure}

To numerically demonstrate the computational properties of the two methods, without loss of generality, we considered simulation scenario S2 in Section \ref{sec:S2} and used $\beta_G = \log(1.5)$ for illustration. Results in Figure \ref{fig:Sim2Figure}(a) confirm our analytical expectation: The computational time of P2.RD grows linearly with respect to $n$, while that of P1.SS is independent of $n$. 

However, the accuracy of P1.SS depends on large $B$; we used $B=$ 10,000 ($\log_{10}(B)=4$) in Figure \ref{fig:Sim2Figure}(a). Figure \ref{fig:Sim2Figure}(b) shows the stability of P1.SS with respect to $\log_{10}(B)$. {For each choice of $\log_{10}(B)$ from 3.0 to 4.3, the $\log_{10}$ sample standard error of the power (SE) computed by P1.SS, obtained from 1,000 independently simulated replicates, was shown in Figure \ref{fig:Sim2Figure}(b). Results clearly show that $B\geq$ 10,000 ($\log_{10}(B)\geq 4$) leads to negligible SE of less than 0.01 ($\log_{10}(\text{SE}) < -2$).} Thus, the default value for $B$ in our method implementation is 10,000. Interestingly, the relationship between $B$ and SE appears to be approximately log-log linear.

We note that Y-axis in Figure \ref{fig:Sim2Figure}(a) measures the run time in seconds for computation of one set of parameter values (i.e.\ per computation). In practice, it is often necessary to run power and sample size analysis for a fine grid of a large number of possible parameter values. Thus, the run time difference aggregates and can differ significantly between P1.SS and P2.RD. In general, when $n$ is larger than 25,000, P1.SS is preferred over P2.RD. Thus, although the implemented software {\tt SPCompute} includes both methods, it sets P1.SS as the default method.

\section{Application to the UK Biobank data}\label{real}

To illustrate the practical utility of the proposed power and sample size computation methods, we applied them to the UK Biobank data \citep{sudlow2015uk, bycroft2017genome}, focusing on the understudied participants with African background. Without loss of generality, we chose hypertension as the binary trait of interest. For completeness, we also analyzed (diastolic) blood pressure, a continuous trait to contrast with the binary trait.

\subsection{Sample and SNP data quality control}\label{sec:QC}

We started with the 3,460 participants with self-reported ancestry being African.  First, we followed the standard practice \citep{marees2018tutorial} to filter out individuals with genotype missingness higher than 20 percent. 
To remove related individuals, we then filtered out individuals with kinship coefficient larger than 0.25, which ended up with 3,182 (approximately) unrelated self-reported Africans. 

\begin{figure}[p]
\centering
\subfigure[All UKB individuals: PC2 vs PC1]{
  \includegraphics[width=0.46\textwidth]{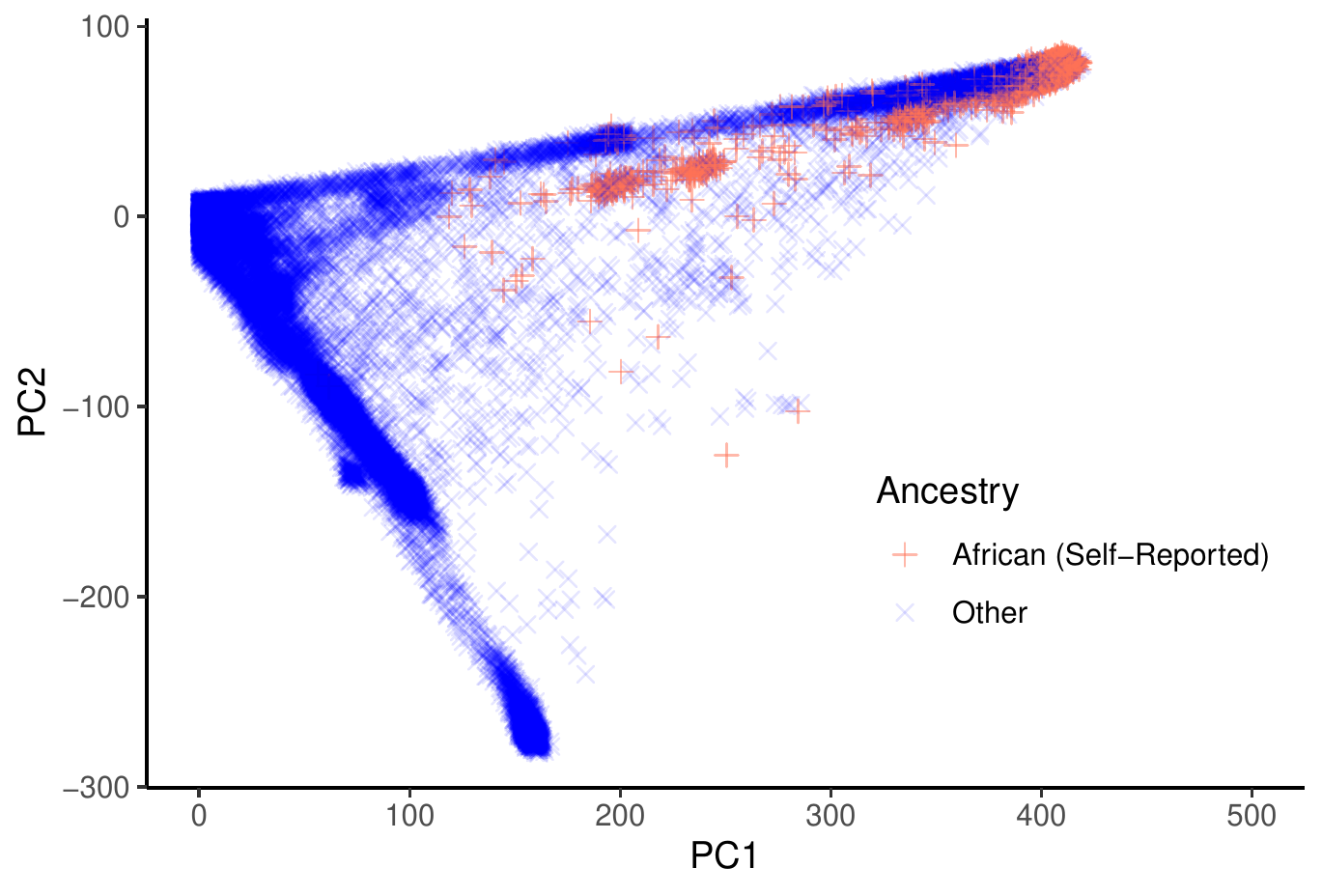}
}
\subfigure[Self-reported Africans: PC2 vs PC1]{
	\includegraphics[width=0.46\textwidth]{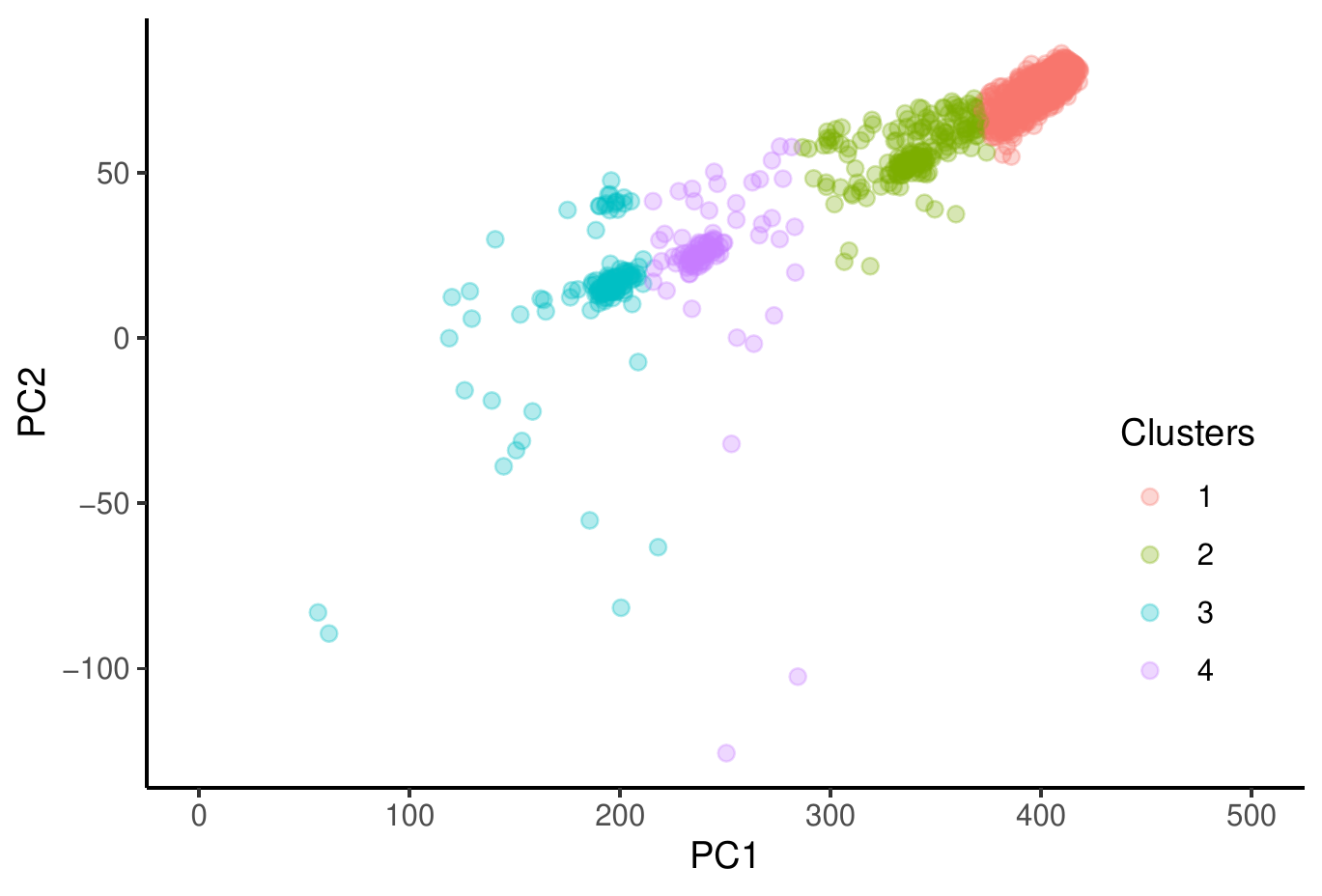}
}
\subfigure[Self-reported Africans: PC3 vs PC1]{
  \includegraphics[width=0.46\textwidth]{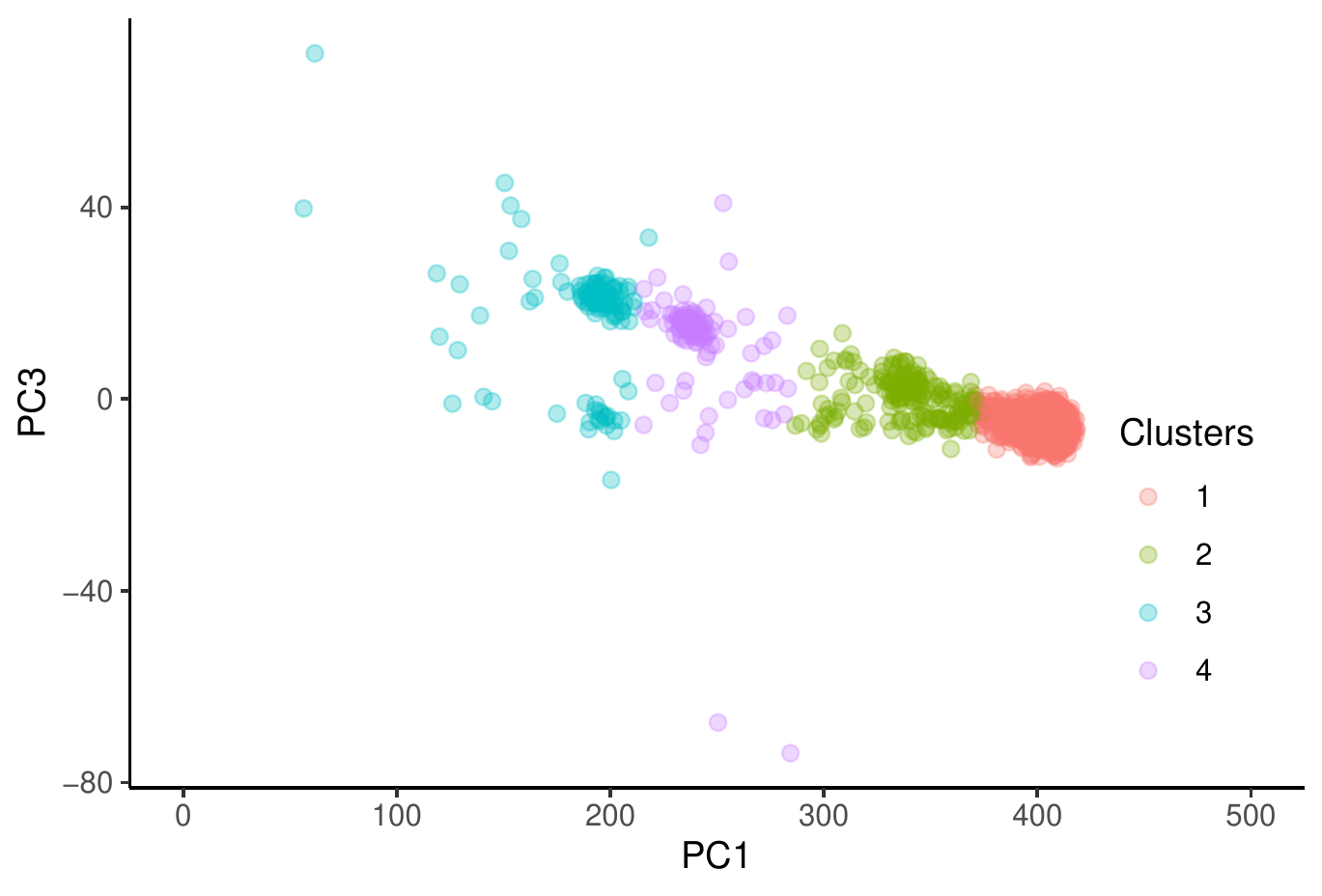}
}
\subfigure[Self-reported Africans: PC3 vs PC2]{
	\includegraphics[width=0.46\textwidth]{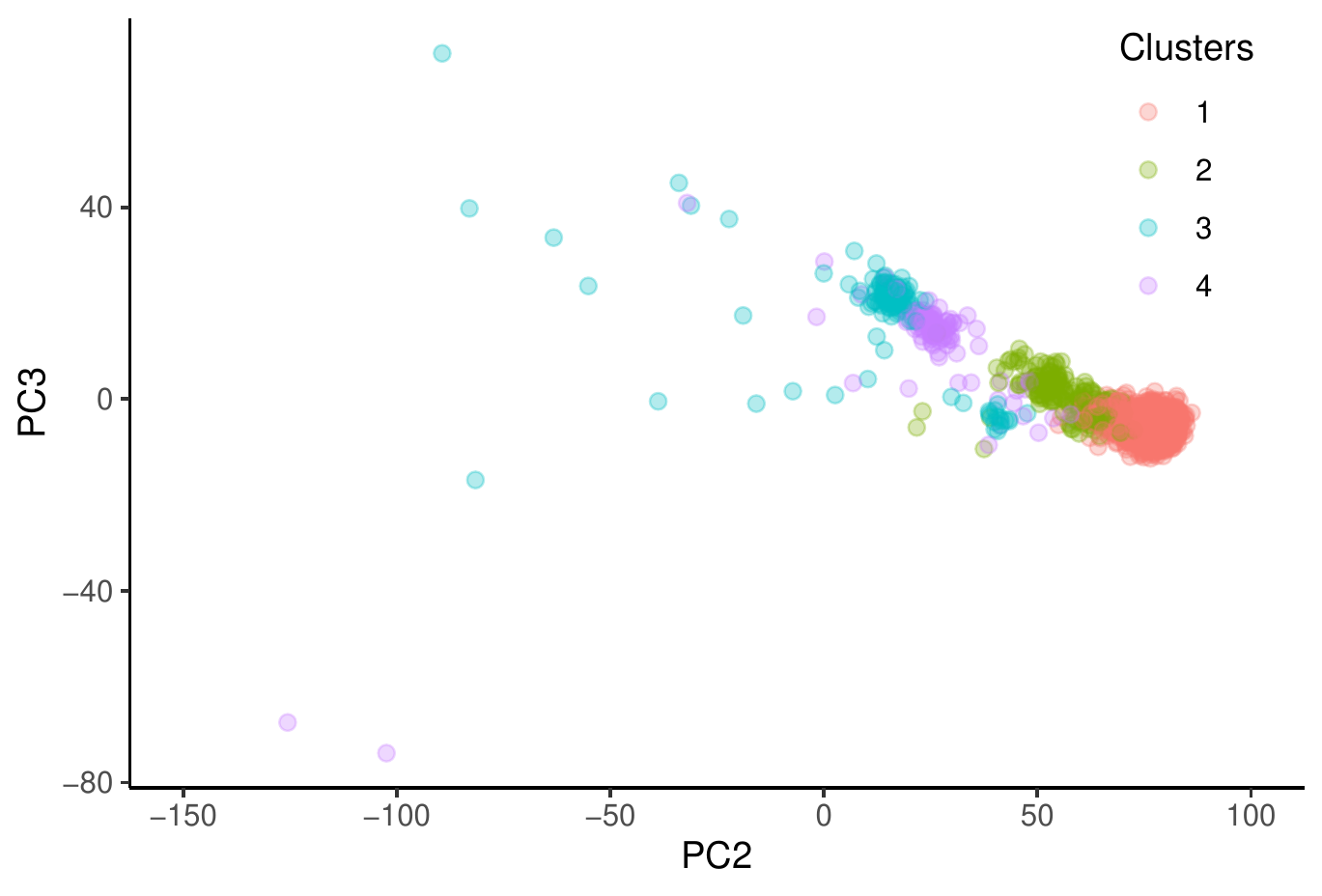}
}
\caption{Population principle components plots for (a) the whole UK Biobank sample, stratified by $n=$ 3,460 self-reported African sample vs.\ Others,  and (b)--(d) the self-reported African sample. In Figures (b)--(d), the four clusters were identified by a K-mean algorithm as discussed in Section \ref{real}. The GWAS shown in Figure \ref{fig:GWAS} used the $n=$ 2,510 individuals in Cluster 1, identified based on this PCA analysis.}
\label{fig:OverallPC}
\end{figure}

To account for reporting error and other ancestry biases, we then performed principle component analysis (PCA) using the overall principle components (PCs) provided by UKB (Data-Field 22009).
Figure \ref{fig:OverallPC}(a) shows the first two PCs of all UKB samples, stratified by self-reported Africans and Others, which suggests reporting error. We then applied a K-mean algorithm with $K = 4$ \citep{hartigan1979algorithm} to the 3,182 unrelated self-reported Africans (Figure \ref{fig:OverallPC}(b-d)) using the overall PCs provided. Cluster 1 contains 2,601 individuals, 75\% of all self-reported African participants (Figure \ref{fig:OverallPC}(b-d)). Following the common practice, we also computed new PCs using only the 3,182 individuals (Supplementary Figure S1). Among the 2,601 individuals in Cluster 1, we then removed 91 individuals whose new PCs were four standard deviations away from the mean. Thus, the final GWAS sample consists of $n=$ 2,510 (1232 females and 1278 males) unrelated individuals with PC-confirmed African ancestry.

For the genetic data, we started with 784,256 genotyped autosomal SNPs (Data-Field 22418), and we then filtered out SNPs based on the thresholds of HWE p-value$<$1e-10, MAF$<0.01$ and missing rate $>$0.2. The X chromosome was not included in our analysis due to the recent report of previously unrecognized data quality issue of the X chromosome \citep{wang2022major}.  
In total, 379,003 common, good quality autosomal SNPs were selected for the subsequent analyses.

\subsection{GWAS of hypertension and blood pressure}\label{sec:GWAS}

We considered two phenotypes, one binary (hypertension) and the other continuous (diastolic blood pressure). In this application, we only considered their measurements at the initial assessment, as longitudinal data analysis is beyond the scope of this work. There are two measurements of blood pressure during the initial assessment, and we used the average. Additionally, for blood pressure we considered the automated reading instead of the manual reading. Among the $n=$ 2,510 analyzed individuals, the prevalence rate of hypertension is 39.48 percent, and the diastolic blood pressure has mean 85.35 and standard deviation 10.75. 

\begin{figure}[p]
\centering
\subfigure[The (binary) hypertension trait]{
  \includegraphics[width=0.46\textwidth]{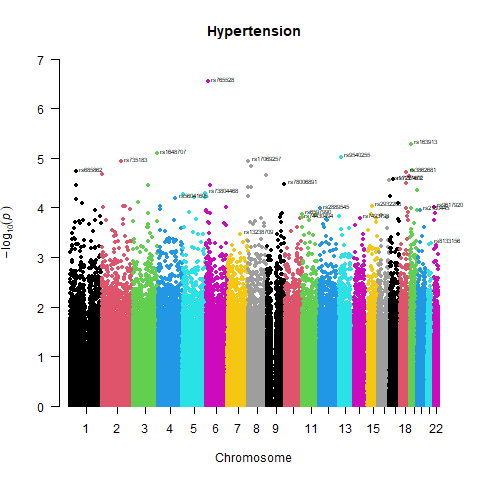}
}
\subfigure[The (continuous) blood pressure trait]{
	\includegraphics[width=0.46\textwidth]{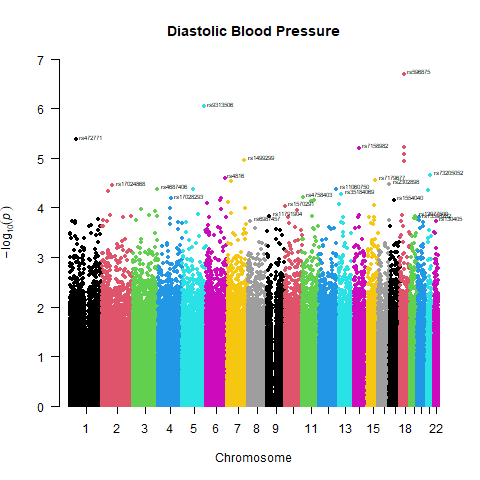}
}
\caption{GWAS Manhattan results for (a) the binary hypertension trait and (b) the continuous diastolic blood pressure trait, using the African sample ($n=$ 2,510) identified through a PCA analysis of the self-identified African sample of the UK Biobank data as discussed in Section \ref{real}. The association analyses included age and sex as important covariates. No SNPs reached genome-wide significance level of 5e-8.}
\label{fig:GWAS}
\end{figure}

The GWAS of both traits included age and sex as covariates. 
The analysis of the binary hypertension trait used logistic regression, and the analysis of continuous trait used linear regression as in convention.  
The two GWAS results are displayed in Figure \ref{fig:GWAS}(a) and (b), respectively, for hypertension and blood pressure. Given the small sample size, it is not surprising that none of the SNPs reached the genome-wide significance of 5e-8 \citep{dudbridge2008estimation}.

\subsection{Power and sample size computation}\label{sec:samplesize}


To illustrate the importance of including covariates in power and sample size computation for a binary trait, as a proof-of-principle, we focused on the top five ranked SNPs from each GWAS. The effect size estimates of both SNPs and covariates were used for the corresponding power and sample size computation, though we recognize the potential issue of winner's curse \citep{sun2011br}; this issue does not change characteristically the conclusion drawn from the methods comparison. 

To illustrate the important role of age and sex in study planning of the binary trait of hypertension, we computed powers and required replication sample sizes twice. We first ignore the two covariates as commonly done (without E), which is equivalent to using {\tt Quanto}; the method of \cite{demidenko2007sample} is not applicable as there are two covariates. We then  accounted for covariate effects (with E) using the proposed method P1.SS, the default method implemented in {\tt SPcompute}. Finally, in addition to $\alpha=0.05$, the standard significance level used for planning a successful replication study, we also considered 5e-8, the genome-wide significance level to demonstrate the power and sample size needed for the {\it discovery} GWAS to achieve 80\% power. For comparison, we also analyzed the continuous trait of blood pressure, where the covariate effects are implicitly incorporated through the specification of the residual variance. 

\begin{figure}[p]
\centering
\subfigure[{\it Overestimated power} for  the (binary) hypertension trait if not accounting for E]{
  \includegraphics[width=0.33\textwidth, height = 0.33\textwidth]{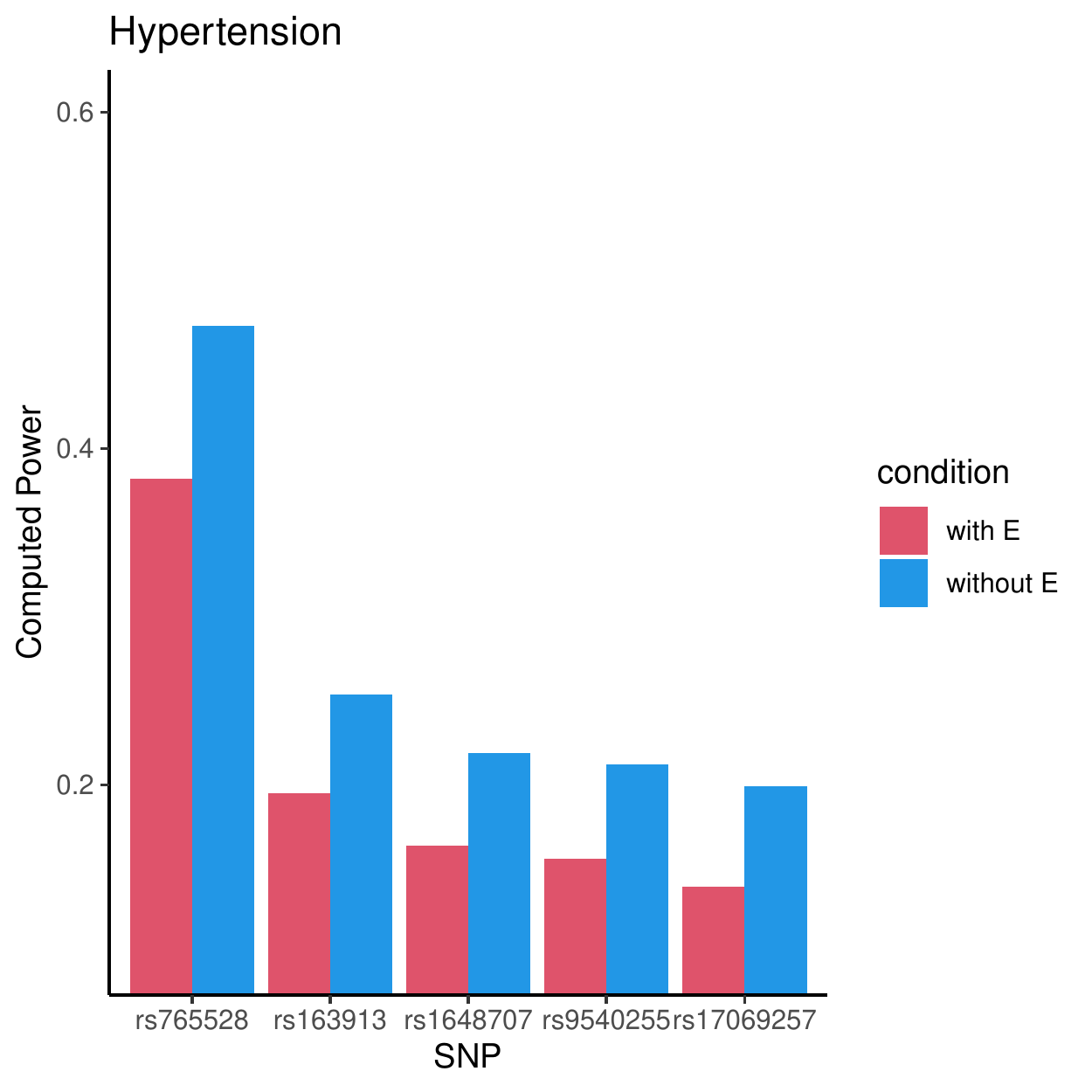}
}
\subfigure[Same power for the (continuous) blood pressure trait]{
	\includegraphics[width=0.33\textwidth, height = 0.33\textwidth]{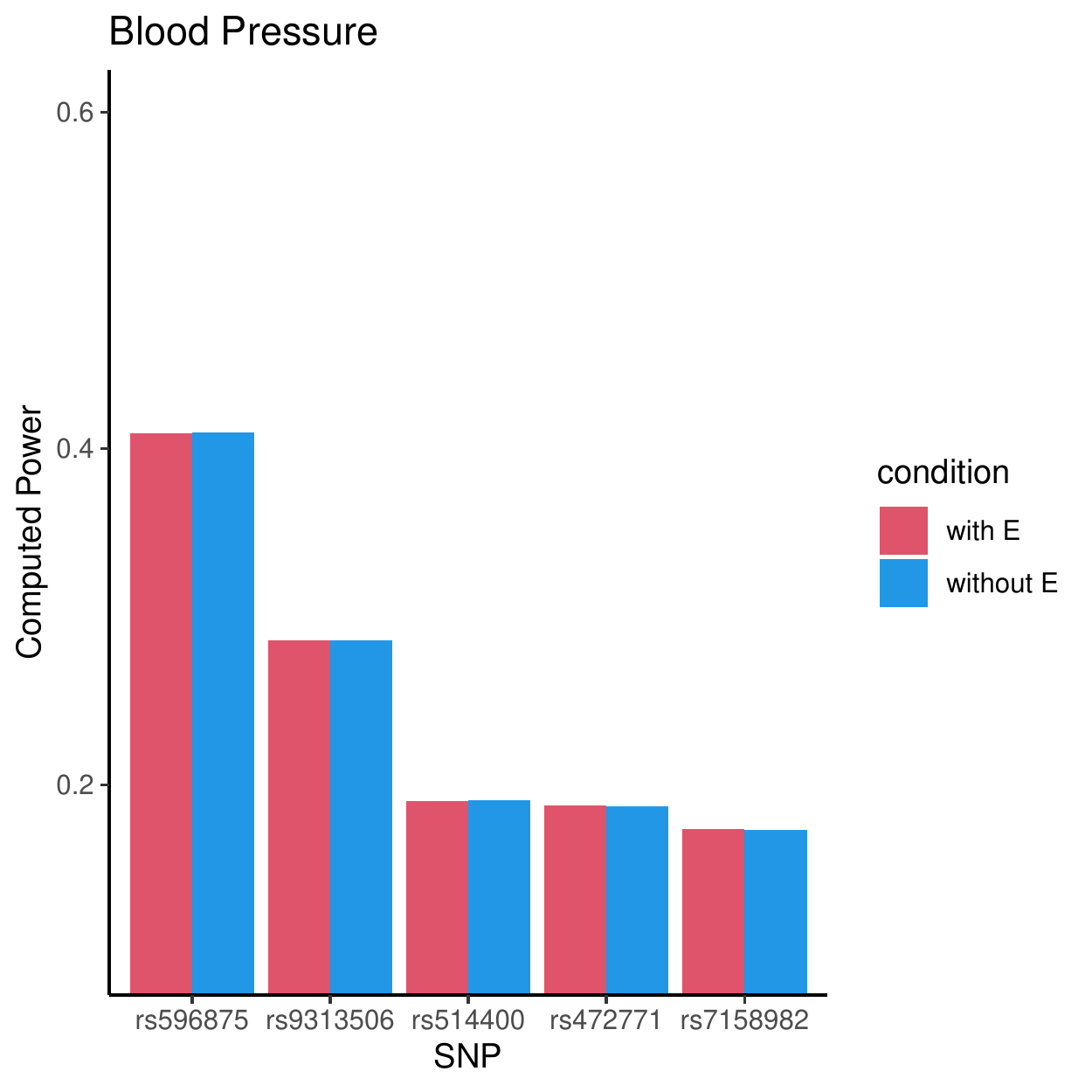}
}

\subfigure[{\it Underestimated (Discovery) sample size} for the (binary) hypertension trait if not accounting for E]{
  \includegraphics[width=0.33\textwidth, height = 0.33\textwidth]{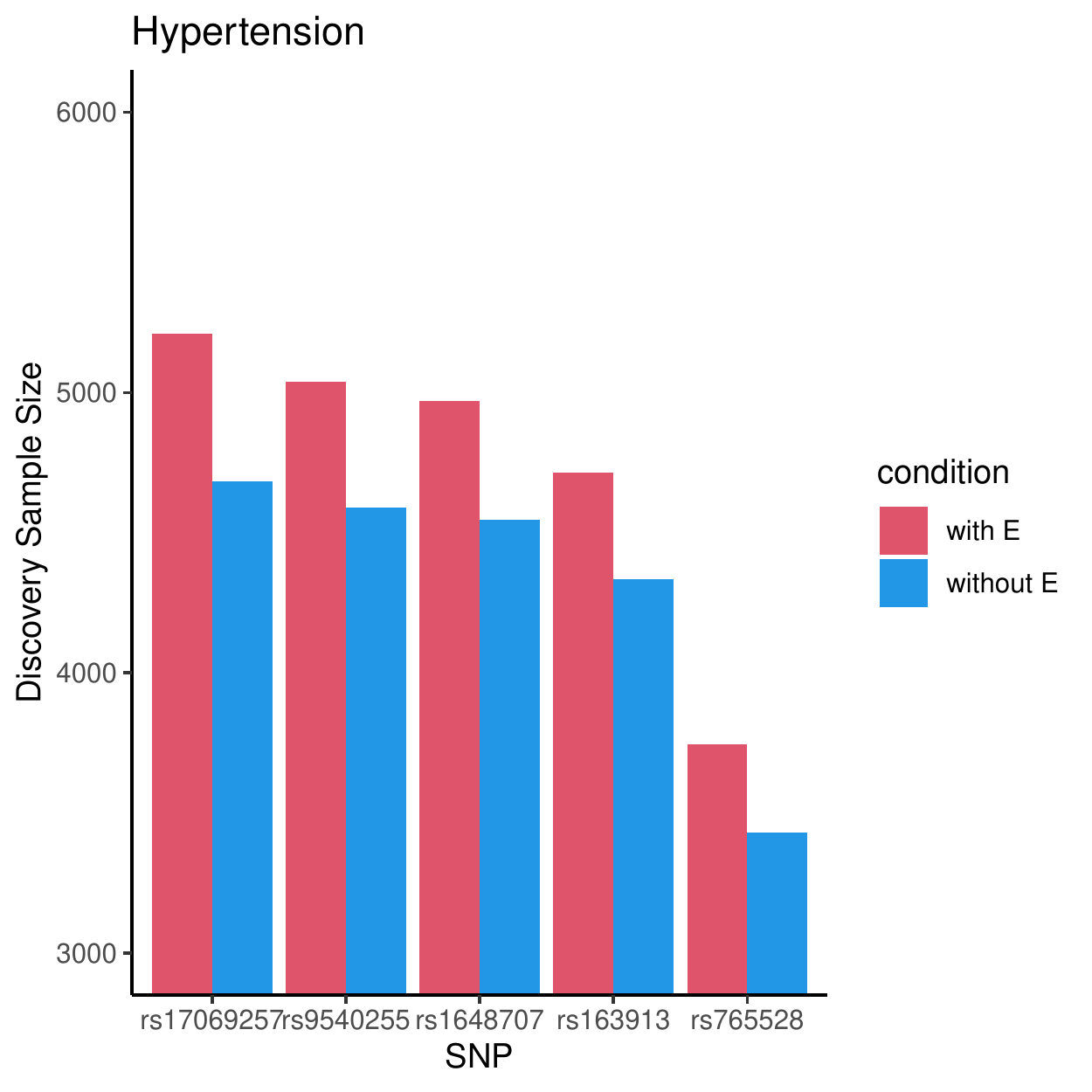}
}
\subfigure[Same (Discovery) sample size for the (continuous) blood pressure trait]{
	\includegraphics[width=0.33\textwidth, height = 0.33\textwidth]{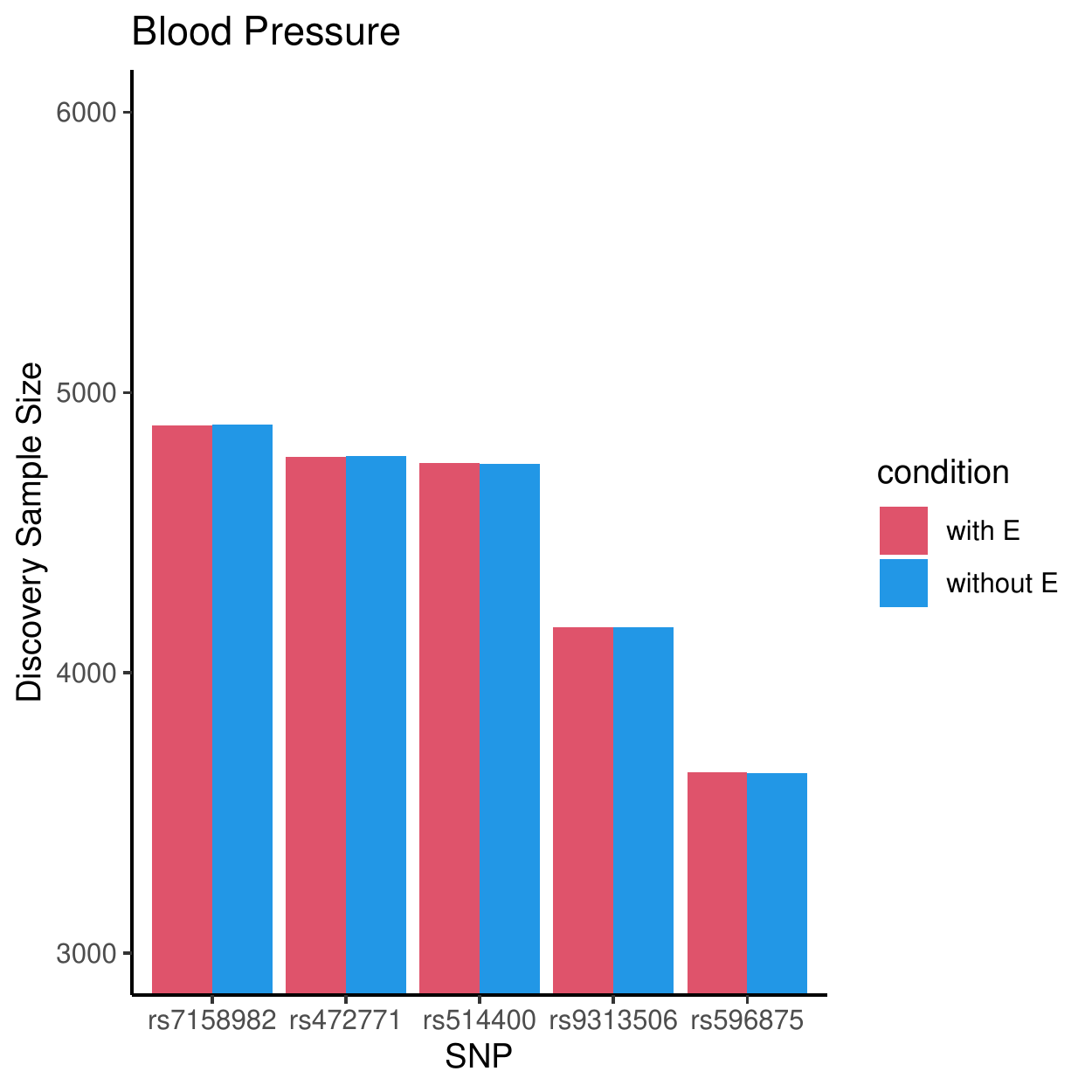}
}

\subfigure[{\it Underestimated (Replication) sample size} for the (binary) hypertension trait if not accounting for E]{
  \includegraphics[width=0.33\textwidth, height = 0.33\textwidth]{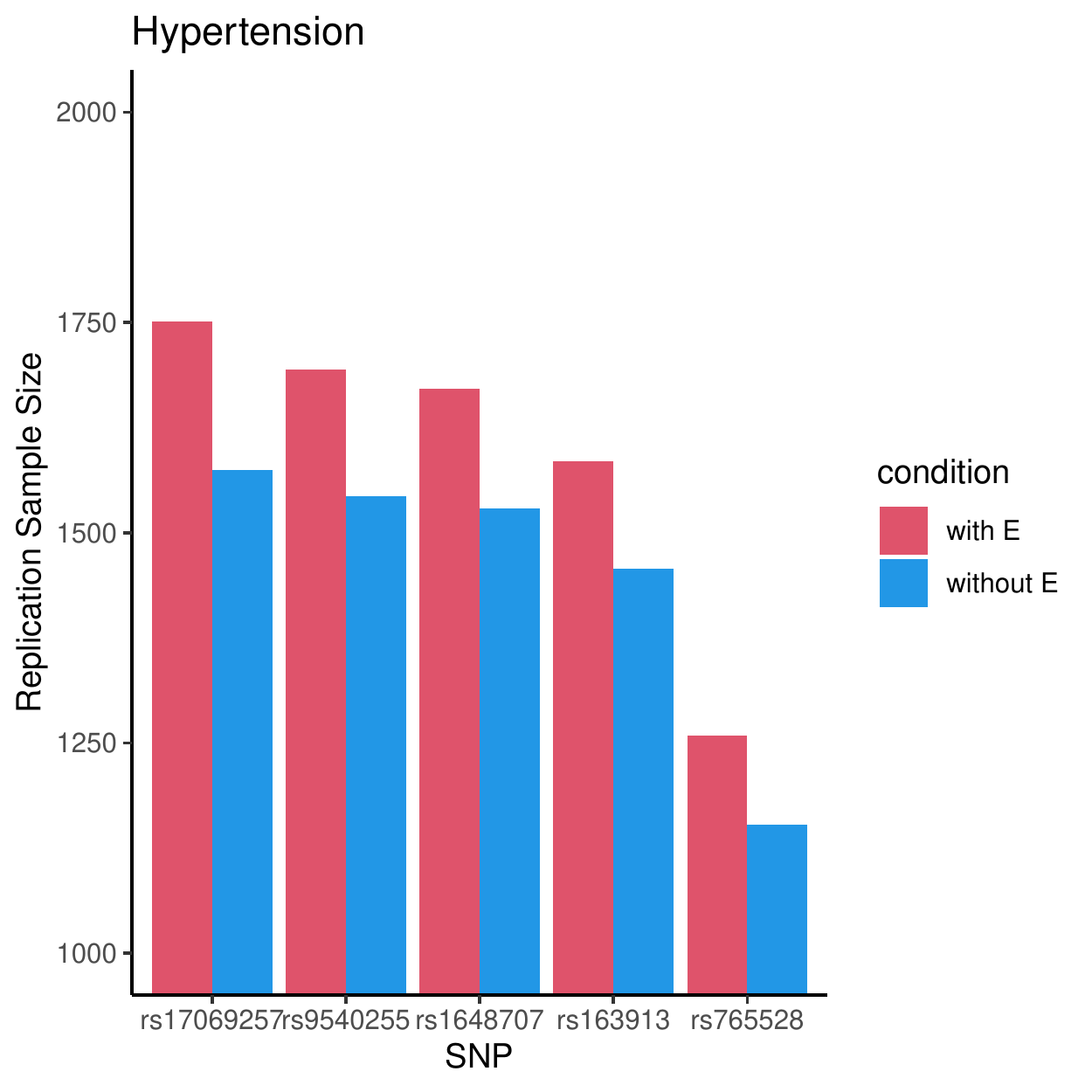}
}
\subfigure[Same (Replication) sample size for the (continuous) blood pressure trait]{
	\includegraphics[width=0.33\textwidth, height = 0.33\textwidth]{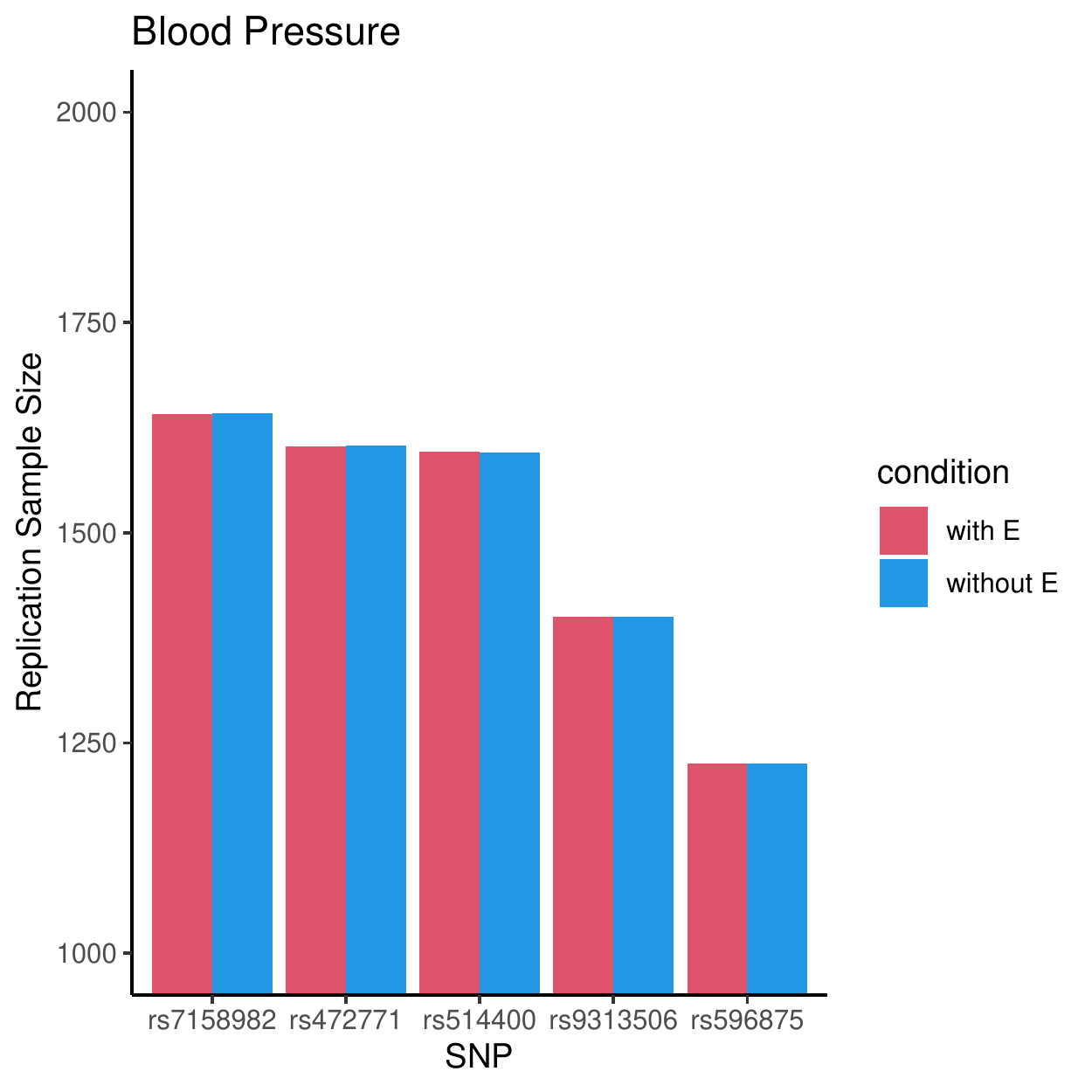}
}

\caption{Powers (Figures (a) and (b)) and sample sizes (Figures (c)-(e)) estimation for study planning of the top five-ranked SNPs identified in GWAS of the binary hypertension trait (Figures (a), (c), (e)) and the continuous diastolic blood pressure trait (Figures (b), (d) and (f)), using the African sample ($n=$ 2,510) identified through a PCA analysis of the self-identified African sample of the UK Biobank data as discussed in Section \ref{real}. The genetic effects of these SNPs used for power and sample size computations are based on a standard GWAS, shown in Figure \ref{fig:GWAS}, where age and sex were included as important covariates. For (replication) study planning, the red bars are the computed power or sample size with adjustment for age and sex, and the blues bars are the values without explicitly considering age and sex. The two approaches do not have difference in power and sample size planning for the continuous blood pressure trait, as age and sex effects are incorporated through residual variance. In contrast, when analyzing a binary trait, the higher blue bars in Figure (a) show that ignoring covariate effects leads to {\it overestimated power} of our discovery study (at $\alpha=$ 5e-8); power for $\alpha=$ 0.05 is close to 100\% as expected, thus not shown. The shorter blue bars in Figure (c) show that ignoring covariate effects leads to {\it underestimated discovery sample size} (for 80\% power at $\alpha=$ 5e-8). Similarly, the shorter blue bars in Figure (e) show that ignoring covariate effects leads to {\it underestimated replication sample size} (for 80\% power at $\alpha=$ 0.05) when planning a replication study of a binary trait.}
\label{fig:TopPower}
\end{figure}

For each trait analyzed, the computed powers and  for the top SNPs are shown in Figure \ref{fig:TopPower}, which are consistent with our simulation results. For the binary hypertension trait (the left panel of Figure \ref{fig:TopPower}), the higher blue bars in Figure (a) show that ignoring covariate effects leads to {\it overestimated power} of our discovery study (at $\alpha=$ 5e-8); the power for $\alpha=$ 0.05 is close to 100\% as expected, thus uninteresting and not shown. The shorter blue bars in Figure (c) show that ignoring covariate effects leads to {\it underestimated discovery sample size} (for 80\% power at $\alpha=$ 5e-8). Similarly, the shorter blue bars in Figure (e) show that ignoring covariate effects leads to {\it underestimated replication sample size} (for 80\% power at $\alpha=$ 0.05) when planning a replication study of a binary trait. In contrast, when studying the continuous blood pressure trait, age and sex effects are already implicitly incorporated through the residual variance. Thus, covariate effects $\beta_E$'s do not have to be explicitly included in the power and sample size computation for a continuous trait.

\section{Discussion}

Adjusting for covariate effect is a standard practice in GWAS, but it is rarely done at the study planning stage and in replication studies. When analyzing a continuous trait through linear regression, covariate effects are implicitly accounted for through residual variance. However, when analyzing a binary trait through logistic regression, covariate effects ($\beta_E$'s) must be explicitly specified and included in power and sample size computation, in addition to the genetic effect of interest ($\beta_G$). 

This phenomenon is known in the statistics literature, but tools available to practitioners are limited. For example, the most well-known software {\tt Qaunto} does not consider $\beta_E$ unless the power analysis is for $G$x$E$ interaction analysis \citep{gauderman2002sample,gauderman2002sample2}, while the method of  \cite{demidenko2007sample} only allows for one binary $E$. In this work, we developed and implemented a flexible software {\tt SPCompute} for accurate and efficient power and sample size computation for a binary trait. We applied the proposed method to the UK Biobank data, analyzing the binary hypertension trait and simultaneously accounting for age and sex covariate effects in power and sample size computation. We also conducted extensive simulation studies to demonstrate the accuracy and efficiency of the proposed method.

However, there are still several limitations of the proposed method that require future work to address. For example, winner's curse where the effect size estimates of significant SNPs are biased upward is known to be a common problem in GWAS \citep{sun2005reduction, zollner2007overcoming,zhong2010correcting,sun2011br}. Therefore, it would be of interest to investigate how {\tt SPCompute} accounts for the winner's curse. Another direction of extension would be to account for mis-classification (particularly the control data), which affects the power of an association study \citep{rekaya2016analysis,zhang2020genetic,lin2020statistical}. 
Additionally, the proposed framework can be further generalized to accommodate the simultaneous analysis of multiple rare variants \citep{derkach2014pooled}. 
Finally, the proposed method assumes a random sample of unrelated individuals. Power and sample size computation for related individuals are worthy future work. 

The proposed method, and the default setting of the implemented software {\tt SPCompute}, assumes all the parameters are specified based on a prospective model as in the UKB application. However, it can also be applied to data collected through the case-control design by modifying the disease prevalence parameter to reflect the case-control ratio used, as shown in \cite{demidenko2007sample}. Additionally, our extensive simulation studies in Section \ref{sim} demonstrated that the proposed method is accurate when parameters were specified based on a prospective model while data were collected through case-control design. However, it should be noted that unlike the regression parameters $\beta_G$ and $\beta_E$, the covariate space $\mathbb{F}_X$ can be very different once being conditioned on the case-control ratio \citep{prentice1979logistic,self1988power}.

Our UKB application in Section \ref{real} only serves as a proof-of-principle and highlights the practical utility of {\tt SPCompute}, such as its ability to handle both binary and continuous covariates. We made some simplifying assumptions to make the example easier to be understood. For example, we accounted for the covariate effects of age and sex simultaneously by introducing two separate models in the second stage regression of (\ref{new-second-stage}). This implicitly assumed the two covariates are conditionally independent given the SNP $G$, an assumption that might be unrealistic in more complex settings. Finally, the framework of the proposed method can be generalized to incorporate the gene-gene and gene-environment interaction effects, which we will provide as future software updates.

%
%

\section*{Acknowledgements}
ZZ is a trainee of the CANSSI-ONTARIO STAGE (Strategic Training for Advanced Genetic Epidemiology) training program at the University of Toronto. 

\section*{Conflict of Interest Statement}
The authors declare that the research was conducted in the absence of any commercial or financial relationships that could be construed as a potential conflict of interest.

\section*{Author Contributions}
ZZ developed the method, performed the analyses, summarized the results, and drafted the manuscript. LS conceptualized and supervised the study. Both authors read and approved the final manuscript.

\section*{Funding}
This research was funded by the Natural Sciences and Engineering Research Council of Canada (NSERC, RGPIN-04934), the Center for Addition and Mental Health (CAMH)
Discovery Fund Seed Funding, and the University of Toronto Data Sciences Institute (DSI) Catalyst Grant. 

\section*{Data Availability Statement}

The UK Biobank data were used under the license for this study (application number {64875}). Data are available at www.ukbiobank.ac.uk/ with the  permission  of UK Biobank. The simulated datasets can be found at the \href{https://github.com/AgueroZZ/Hidden_Factor_Code}{online repository} of this manuscript.

%
%

\nocite{*}
\bibliographystyle{natbib}
\bibliography{myrefs}%

\clearpage

\begin{table}[p]
  \begin{center}
  \scalebox{1}{
  \begin{tabular}
  {|p{1.5cm}|p{1.2cm}p{1.2cm}|p{1.2cm}p{1.2cm}|p{1.2cm}p{1.2cm}|}
  \hline
  \cline{1-7}
        \multicolumn{1}{l}{} &
        \multicolumn{2}{c}{Scenario 1 (S1) } &
        \multicolumn{2}{c}{Scenario 2 (S2)} &
        \multicolumn{2}{c}{Scenario 3 (S3)} \\
        \hline
        Methods &  {Average AE} & {Maximum AE} & {Average AE} & {Maximum AE} &  {Average AE} & {Maximum AE} \\
   \hline
   P1.SS & 0.008 & 0.022 & 0.010 & 0.021 & 0.015 & 0.032  \\
   P2.RD & 0.009 & 0.019  & 0.012  & 0.023  & 0.012 & 0.027 \\
   Demidenko & 0.009  & 0.019 & 0.012 & 0.024 &  0.124  & 0.194 \\
     &  &  & &  &  0.097*  & 0.150*\\
    {Quanto} & 0.009  & 0.022 & 0.052 & 0.084 & 0.112  & 0.173 \\
  \hline
  \end{tabular}
  }
  \end{center}
  \caption{The average and maximum absolute error (AE), across different sample sizes, between the oracle and computed power using the different methods for the three scenarios considered. P1.SS is the proposed  `semi-simulation' method in Section \ref{m1}, and P2.RD is the proposed `representative dataset' method in Section \ref{m2}.  In Scenario 3, the method of \cite{demidenko2007sample} was implemented by dichotomizing $E$ or without considering $E$ (results*). See legend to Figure \ref{fig:Sim1Figures} for additional details.}
  \label{table:Sim1table}
\end{table}

\end{document}